\documentclass[prb,superscriptaddress,showpacs,twocolumn]{revtex4}

\usepackage{amsmath,amssymb,graphicx,color,bm,soul,ulem}
\usepackage{hyperref}

\begin{document}

\title{Quantum statistics of light emitted from a pillar microcavity}

\author{T.~A.~Khudaiberganov}
\affiliation{Department of Physics and Applied Mathematics, Vladimir State University named after A. G. and N. G. Stoletovs, 87 Gorkii st., 600000 Vladimir, Russia}

\author{S.~M.~Arakelian}
\affiliation{Department of Physics and Applied Mathematics, Vladimir State University named after A.~G. and N.~G. Stoletovs, 87 Gorkii st., 600000 Vladimir, Russia}

\date{\today}

\begin{abstract}
A quantum behavior of the light emitted by exciton polaritons excited in a pillar semiconductor microcavity with embedded quantum well is investigated. Considering the bare excitons and photon modes as coupled quantum oscillators allows for an accurate accounting of the nonlinear and dissipative effects. In particular, using the method of quantum states presentation in a quantum phase space via quasiprobability functions (namely, a $P$-function and a Wigner function), we study the effect of the laser and the exciton-photon detuning on the second order correlation function of the emitted photons.
We determine the conditions for the phenomena of bunching, giant bunching, and antibunching of the emitted light. In particular, we predict the effect of a giant bunching for the case of a large exciton to photon population ratio. Within the domain of parameters supporting a bistability regime  we demonstrate the effect of  bunching of photons.
\end{abstract}

\maketitle

\section{Introduction}
Exciton polaritons are mixed quasiparticles arising due to the strong coupling of photonic mode with an exciton resonance. Being initially considered as a fundamental example of composite bosons with light effective mass, which is attractive for the high-temperature condensation, the exciton polaritons are recognized now to be well suited for realization of the practical devices being competitive with the state of the art optoelectronic and photonic devices. Indeed, the composite nature of exciton polaritons takes advantage of both photonic and excitonic constituents. Namely, polaritons inherit a high mobility and the ease of excitation from photons as well as the strong two-body interactions from excitons. This combination makes polaritonic systems a versatile platform for studying  quantum and nonlinear phenomena in strongly coupled light-matter systems.

We consider the exciton polaritons (here after polaritons) formed in a pillar microcavity (or abbreviation is micropillar) \cite{GBM} -- see the sketch shown in Fig.~1. The micropillar has a narrow optical cone. Therefore, in contrast to a planar microcavity, it supports formation of zero-dimensional polaritons whose spatial degrees of freedom are suppressed. In other words, the in-plan wave vector of polaritons $\vec{k}$ is normal to the growth axis of the structure $k=0$.

The quantum and nonlinear properties of exciton polaritons is of a great interest. When the cavity is driven by the external laser field, the nonlinear behavior can be revealed in a bistable optical response~\cite{BKG} of the medium. This  behavior is classical and can be observed by the sudden jumps of a transmission characteristic during pump intensity scanning. Here, in contrast, we are interested in the quantum manifestations of the bistability effect, i.e., in the statistics of the light emitted by the microcavity. In particular, we are aimed at searching of the conditions when the system behavior will be much different from what  the classical description of the problem predicts.For example, the influence of a classical noise on the exciton-polariton bistability has already been considered in the article by Abbaspour et al.~\cite{AST}.
At the same time, in the experimental works \cite{RCS,FSH}, a bistable response characterized by the hysteresis loop was observed to be narrower than it was expected in the classical case. Such a squeezing of the bistability loop indicates on a significant impact of the quantum noise effects on the polariton behavior.

A general characteristic of the quantum properties of a given system is a second order correlation function $g^{\left(2\right)}(\tau) =\frac{\left\langle a(t) ^{+} a(t+\tau) ^{+} a(t) a(t+\tau) \right\rangle }{\left\langle a(t) ^{+} a(t) \right\rangle ^{2} } $ or in zero delay $\tau=0$ is $g^{\left(2\right)}(0)=1+\frac{\left\langle (\Delta n)^{2} \right\rangle-\left\langle n \right\rangle}{\left\langle n \right\rangle^{2}}$. In particular this function was considered as an important criterion for demonstrating Bose condensation in the exciton-polariton system formed in a planar microcavity \cite{DWS}. The second order correlation function characterizes such quantum statistical properties of radiation as bunching ($g^{\left(2\right)}>1$) and antibunching ($g^{\left(2\right)}<1$). The effect of antibunching was studied for polaritons formed in a quantum box in the context of so-called polariton blockade \cite{VCC}, when the emission of photons in pairs is suppressed and photons are emitted individually. Recent experimental work \cite{KFA} reports on the observation of a sharp peak of the second order correlation function of photons emitted from the micropillar driven by the laser light. However in a planar microcavity a smooth decrease of the second order correlation function above the Bose-Einstein condensation threshold was detected \cite{KFA}. The quantum behavior of excitons in the exciton polaritons systems were investigated in \cite{DKC}. In particular, the peak of the second order correlation function for excitons was obtained. In this article, we focus on the investigation of the quantum behavior of the photonic mode.
%

\begin{figure}
\includegraphics[width=1\linewidth]{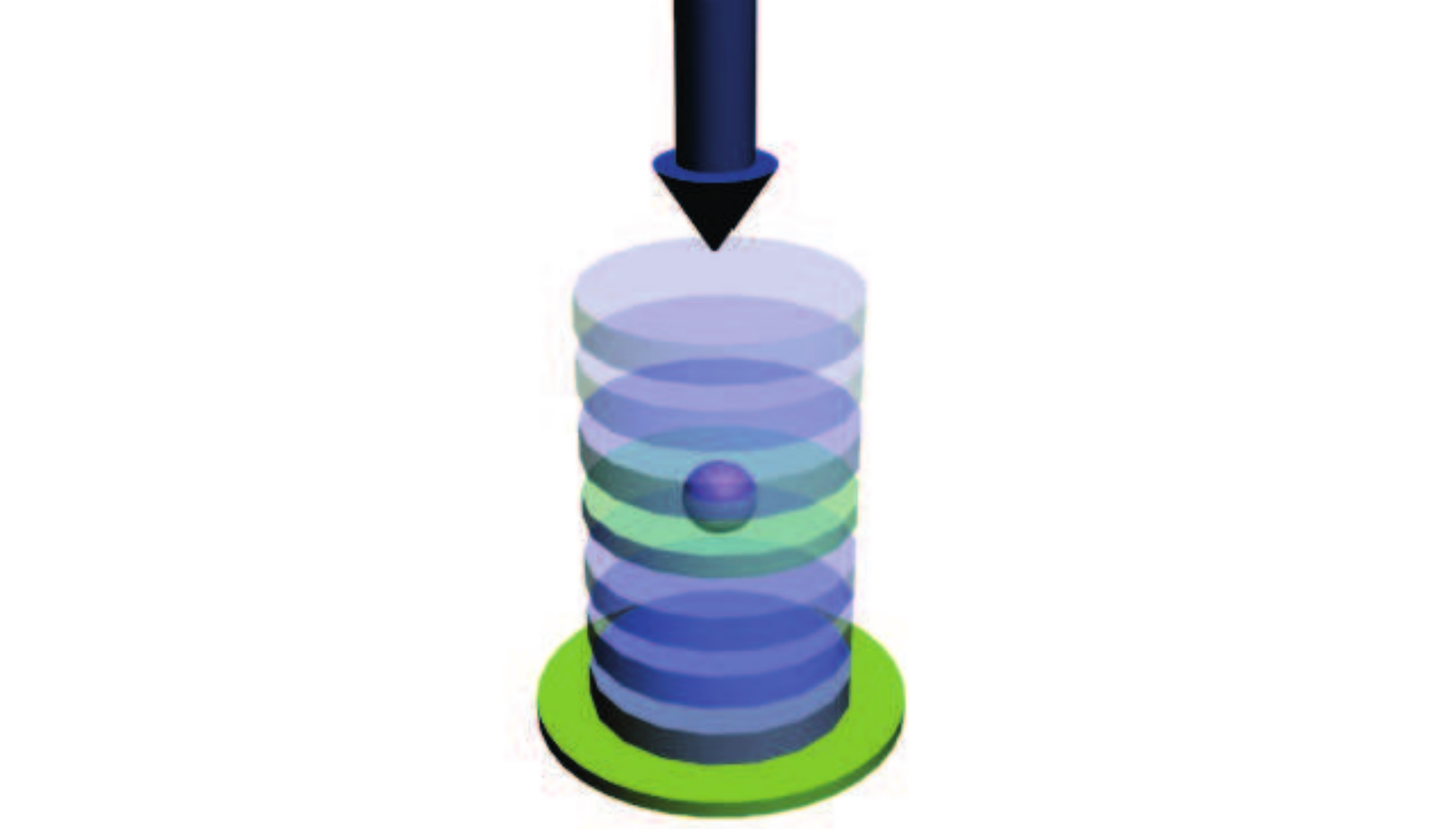}
\caption{~A sketch of the micropillar with embedded quantum well where the zero-dimensional polaritons (purple dot) are formed. The arrow illustrates the external laser pump which drives the photonic mode of the cavity.}
\label{Fig.figure1}
\end{figure}

The paper is organized as follows. In Sec.~1 we introduce the Hamiltonian of the system and derive the master equation. Next, we move  to the description of this system in a quantum phase space, namely, to the $P$-representation. We solve the Fokker-Planck equation for the $P$ function using the  method of potential \cite{DGG} and obtain the first and second order correlation functions \cite{DKC} of the photons using the governed principle. Note that the perturbation theory does not predict the observable peak of the second order correlation function in this case. In Sec.~2, we investigate the effect of quantum fluctuations on the bistability. In Sec.~3, we study the effects of the giant bunching and antibunching effects in the region  of a triple resonance where the exciton, photon and the laser frequencies are close to each other, $\omega _{ex} \approx \omega _{ph} \approx \omega _{d}$.
\section{The model}
Polaritons are excited by a coherent laser pump which frequency is close both to the photon $\omega _{ph} $ and exciton $\omega _{ex} $ frequencies. The Hamiltonian of two coupled modes with the coherent pumping in the rotating wave approximation \cite{BKG,DKC} reads:
\begin{equation} \label{Eq1_}
\begin{array}{l} {\hat{H}_{S}=\hbar \Delta _{{\rm ph}} \hat{\phi }^{+} \hat{\phi }-\hbar \Delta _{{\rm ex}} \hat{\chi }^{+} \hat{\chi }+}\\{+\hbar \omega _{{\rm R}} \left(\hat{\chi }^{+} \hat{\phi }+\hat{\phi }^{+} \hat{\chi }\right)+\hbar \alpha \hat{\chi }^{+2} \hat{\chi }^{2}+{ i}\hbar \left(\tilde{E}_{d}\hat{\phi }^{+} -\tilde{E}_{d}^{*}\hat{\phi }\right).} \end{array}
\end{equation}
Here $\hat{\phi }$ ($\hat{\phi }^{+} $) and $\hat{\chi }$ ($\hat{\chi }^{+} $) are annihilation (creation) operators of the photon and the exciton modes respectively.   $E_{d}=\tilde{E}_{d}e^{-i\omega_{d}t}$ -- is an external pump strength, where $\omega_{d}$ -- is a frequency of the coherent pump. The value $I_d=|E_{d}|^{2}$ is proportional to the intensity of the laser pump.
The following detunings  were introduced: $\Delta _{ph} =\omega _{ph} -\omega _{d}$ stands for the detuning of the photon frequency from the frequency of the pump, $\Delta _{ex} =\omega _{ex} -\omega _{d}$ -- is the  detuning of the excitonic frequency from the frequency of the pump, $\omega _{{\rm R}} $ is a half-part of the Rabi splitting, $\alpha $ -- is a coefficient of a Kerr-like nonlinearity. 

The polariton condensate under consideration is an open quantum system  which is inevitably affected by the presence of noise. We suppose that the micropillar temperature is around several Kelvin. In this case, the  thermal noise is weak since the influence of the heat reservoir decreases exponentially with increasing ratio $\hbar \omega _{{\rm ph,ex}} /k_{B} T$. In our case $\hbar \omega _{ph,ex} \gg k_{B} T$, therefore we neglect by the effect of thermal noise. Also, we exclude the noise from the driving intensity \cite{AST}. Thus we are left with a purely quantum noise whose impact on the system behavior is studied below.

We treat the losses of the exciton-photon system via the Lindblad master equation (in the Born-Markov approximation) for the density operator $\rho $ \cite{DKC},
\begin{equation} \label{Eq2_} 
\begin{array}{l} {\frac{\partial \rho }{\partial t} =\frac{1}{{i}\hbar } \left[\hat{H}_{S},\rho \right]+\gamma _{{\rm ph}} \left(2\hat{\phi }\rho \hat{\phi }^{+} -\rho \hat{\phi }^{+} \hat{\phi }-\hat{\phi }^{+} \hat{\phi }\rho \right)+} \\ {+\gamma _{{\rm ex}} \left(2\hat{\chi }\rho \hat{\chi }^{+} -\rho \hat{\chi }^{+} \hat{\chi }-\hat{\chi }^{+} \hat{\chi }\rho \right),} \end{array}
\end{equation} 
where $\gamma _{{\rm ph}} $ and $\gamma _{{\rm ex}} $ are the damping rates of photonic and excitonic modes respectively.

The semiclassical solution of the problem has been described in a number of works \cite{BKG,DKC,YEL}. In particular, it predicts the bistable optical response of the microcavity which manifests itself in a hysteretic behavior of the output light intensity \cite{BKG} -- see Fig.~2b. The optical bistability effect occurs within the particular range of the driving field intensities which is determined by a competition between the losses and the positive nonlinear feedback of microcavity excitons \cite{AST}.

The steady state quantum solution of master equation \eqref{Eq2_} is obtained analytically using the $P$-function approach -- see Appendix A. In what follows we assume that the quantum fluctuations of photons are governed by the quantum fluctuations of excitons \cite{G,H2013} according to the Haken's slaving principle. Thus using the first equation from (A4), in the steady state regime we obtain the following relation between the amplitudes of the photon and the exciton fields:
\begin{equation} \label{Eq3_} 
\phi =\frac{\tilde{E}_{{\rm d}} \left(\gamma _{{\rm ph}} -{\rm i}\Delta_{ph} \right)}{\left(\Delta_{ph}^{2} +\gamma _{{\rm ph}}^{2} \right)} -\frac{\omega _{R} \left({\rm i}\gamma _{{\rm ph}} +\Delta_{ph}\right)}{\left(\Delta_{ph}^{2} +\gamma _{{\rm ph}}^{2} \right)} \chi.                                       
\end{equation} 

In this case a first order correlation function of photons, which is equivalent to a quantum average value of the photon number, is defined as
\begin{equation} \label{Eq4_}
\left\langle \phi ^{+} \phi \right\rangle =\frac{I_{d} +\omega _{R}^{2} G^{\left(11\right)} +2\omega_{R}\textsf{Im}\left(\tilde{E}_{d}G^{\left(10\right)} \right)}{\left(\Delta _{ph}^{2} +\gamma _{ph}^{2} \right)}.
\end{equation}

Here $G^{\left(ij\right)}=\left\langle \left(\chi^{+}\right)^{i}\chi^{j}\right\rangle$ -- is a correlation function of the exciton mode, see Appendix A.

We characterize the quantum properties of the emitted light with the second order correlation function $g_{ph}^{\left(2\right)} =\frac{\left\langle \phi ^{+2} \phi ^{2} \right\rangle }{\left\langle \phi ^{+} \phi \right\rangle ^{2} } $. From \eqref{Eq3_} with the use of Eq.~(A.3) given in the Appendix A one obtains: 
\begin{equation} \label{Eq5_}
\begin{array}{l} {g_{ph}^{2}(0)=} \\  {\frac{\left[I_{d}^{2} - \omega _{R}^{2} (\tilde{E}_{d}^{2}G^{\left(20\right)}+\tilde{E}_{d}^{*2} G^{\left(02\right)}) +2i\omega _{R} I_{d} (\tilde{E}_{d}G^{\left(10\right)}-\tilde{E}_{d}^{*} G^{\left(01\right)}) \right]}{I_{d} +\omega _{R}^{2} G^{\left(11\right)} +i\sqrt{I_{d}}\omega_{R}\left(G^{\left(10\right)} -G^{\left(01\right)} \right)} +} \\ {\frac{\left[2i\omega _{R}^{3} \left(\tilde{E}_{d}G^{\left(21\right)} -\tilde{E}_{d}^{*} G^{\left(12\right)} \right)+4\omega _{R}^{2} I_{d} G^{\left(11\right)} +\omega _{R}^{4} G^{\left(22\right)} \right]}{I_{d} +\omega _{R}^{2} G^{\left(11\right)} +i\omega_{R}\left(\tilde{E}_{d}G^{\left(10\right)} -\tilde{E}_{d}^{*} G^{\left(01\right)} \right)} .} \end{array}
\end{equation} 

\section{The effect of quantum fluctuations on the bistability}

The steady-state state solution of the problem can be obtained in the mean field approximation by averaging  quantum operators $\left\langle \hat{O}\right\rangle _{} =Tr\left(\rho _{} \hat{O}\right)$ and assuming factorization of quantum averages as $\left\langle \hat{\chi}^{+}\hat{\chi}\hat{\chi}\right\rangle\rightarrow|\chi_{mf}|^{2}\chi_{mf}$. In this case the bistability is manifested in the typical $S$-shape  intensity-dependence of the photon mode population, see the dashed curves in Fig.~2b. The region where the solution is bistable, is displayed on the parameter plane of $\Delta $ and $\Omega $ parameters, Fig.~2a. Here $\Delta =\frac{\omega _{ph} -\omega _{ex} }{2} $ and $\Omega =\omega _{{\rm d}} -{\left(\omega _{{\rm ph}} +\omega _{{\rm ex}} \right)\mathord{\left/ {\vphantom {\left(\omega _{{\rm ph}} +\omega _{{\rm ex}} \right) 2}} \right. \kern-\nulldelimiterspace} 2} $ is a detuning of the driving field from the central frequency between the exciton and the photon resonances. The false colors in Fig.~2a correspond to the critical intensity of the driving field above which the single-valued solution is superposed by a bistable one. This value is refereed to as a bistability threshold. There two bistability domains: The upper and the lower one. Both domains are located above the relevant polariton branches $\Delta _{LP,UP} =\pm \sqrt{\omega_{R}^{2} +\Delta ^{2} } $ shown with the blue lines. In particular, the regions of bistability are shifted upwards in frequency on the value which is determined by the level of losses \cite{BKG}.

The upper and lower solutions of the bistability curve shown in Fig.~2b are stable in the mean field approximation. This means that polaritons excited in these states live infinitely long. 
However in the presence of a sufficient external noise the stochastic switching between bistable states becomes possible \cite{CS}.
The problem of stochastic switching in the exciton-polariton system has already been considered in the context of a dissipative phase transition \cite{FSH}. A study of the micropillar radiation demonstrated the stochastic switching between the bistable states  \cite{FSH,KFA}. Besides, the non-classical behavior was discovered in the quantum statistics of the micropillar radiation \cite{FSH}. The early theoretical works \cite{FSH,KFA,CS} which address this question used a truncated polariton approximation \cite{BKG}. It means that they took into account only the lower polariton state neglecting by all other terms in the Hamiltonian \eqref{Eq1_} written in the polariton basis. This approximation is valid when the pump frequency is tuned close to the resonance of the lower polariton branch. However, in the general case of arbitrary frequency of the driving, one needs to consider the interaction with both the upper and lower polaritons. In contrast to the previous studies, here we consider a complete model operating in the exciton-photon basis using the slaving principle introduced by Haken \cite{G} and obtained analytical solution. This approach allows for expanding of the parameters region covered by the quantum nonlinear effects caused by the bistability effect. It is also worth noting that, was done within the framework of numerical simulation of a two-mode system in the work \cite{VCC} for small value of nonlinearity.


As opposite to the mean-field solution predicting two stable stationary states within the bistability region, the quantum approach  always predicts \cite{DGG} a single-valued solution \eqref{Eq4_}, see the solid curve in Fig.2b. However because of the intrinsic quantum noise, the bistable behavior still can be observed within the quantum approach by the presence of the hysteresis loop. The quantum approach treats the classical upper and lower states of the bistability curve as metastable. Whether the system jumps to the different metastable state or not depends on the time spent in the initial state. Therefore, if one scans the driving intensity up and down, the classical hysteresis loop is revealed only in the case of quasi-adiabatic variation of the pump power. At a finite scanning velocity the dynamical hysteresis loop  becomes narrower as it was demonstrated in \cite{RCS}. The loop width depends on the pumping rate increment and on the metastable states lifetime \cite{RCS} (is called a dynamic hysteresis loop). 

The Fokker-Planck equation is associated with a certain stochastic process, which can be described by the stochastic equations (A.4) \cite{DGG,DMW}. Thus, in accordance with the approach of quantum phase spaces, one can describe the dynamics of the quantum variables (c-numbers $\phi,\phi^{+},\chi,\chi^{+}$ is commutative variables) using stochastic differential equations (A.4) with the diffusion and drift taken from the Fokker-Planck equation (A.1). The details are given in the Appendix A. 

In Fig.~2b, solid curves show predictions of the quantum solutions for various laser detunings (indicated by the dots in Fig.~2c). 
We observed the presence of quantum jumps between two metastable states in the pump intensity region where the mean photon number predicted by the quantum solution rapidly growth from low bistable state to the upper one. The stochastic dynamics demonstrating the quantum jumps is shown in Fig.~2e. The parameters of the system correspond to the blue curve in Fig.~2b. The metastable states are indicated by points $2$ and $2^\prime$ connected with the red arrow. The stable states appear in the region where the quantum solution is close to the semiclassical one (points 1 and 3 on the blue curve and the point 4 on the green curve). Note that for the case of a wide bistability loop, there is typically only one stable classical solution branch while the states on the opposite branch of loop are metastable. This case is illustrated by the point 7  which  corresponds to the metastable state located at the lower branch of the green curve. The stochastic dynamics of the system initially excited in the the state 7, as shown in Fig.~2f, demonstrates an irreversible jump to the state $7^\prime$  corresponding to stable upper state. 

\begin{figure}
\begin{minipage}[h]{0.99\linewidth}
\center{\includegraphics[width=1\linewidth]{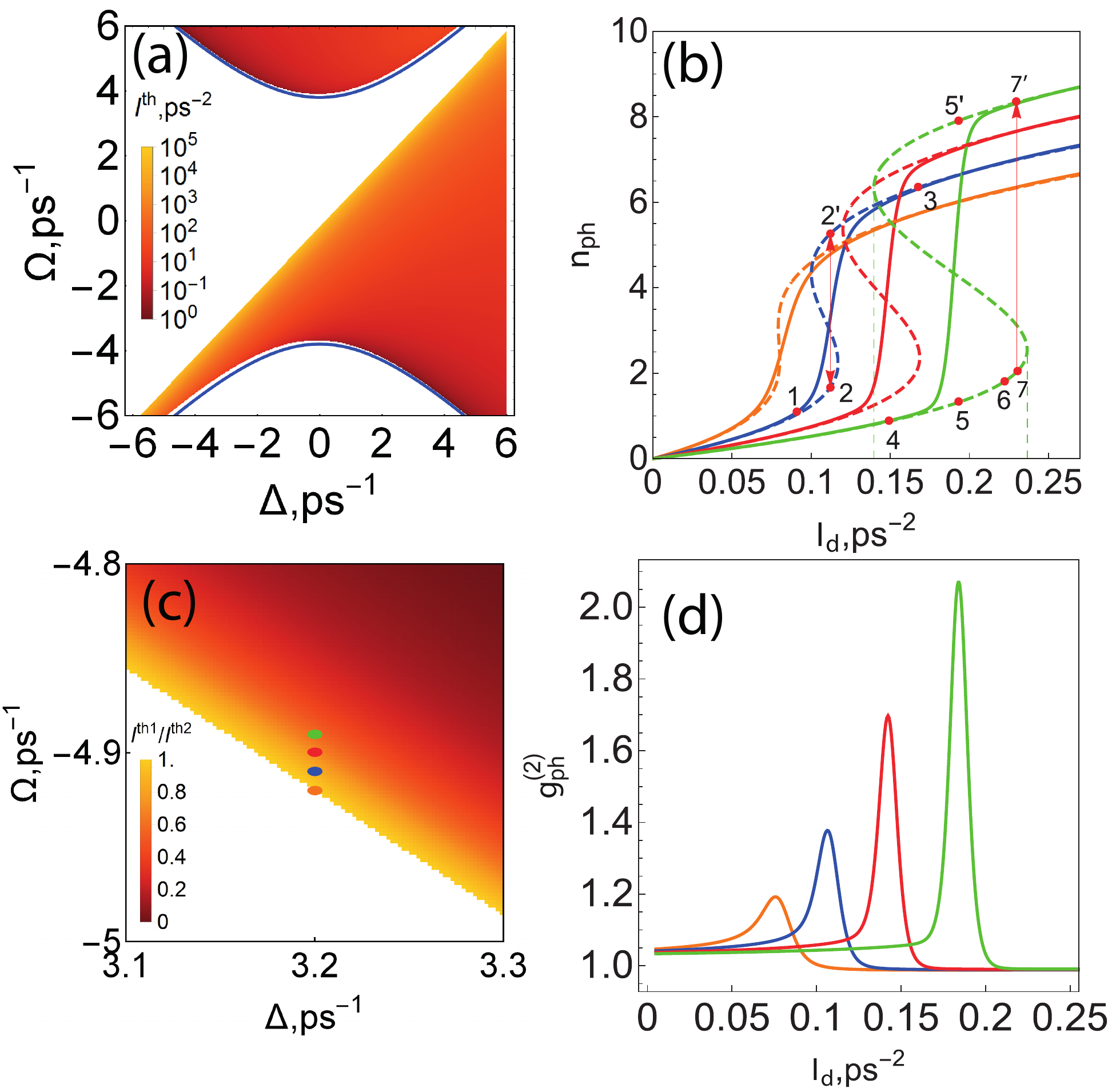}}  
\end{minipage}
\vfill
\begin{minipage}[h]{0.99\linewidth}
\center{\includegraphics[width=1\linewidth]{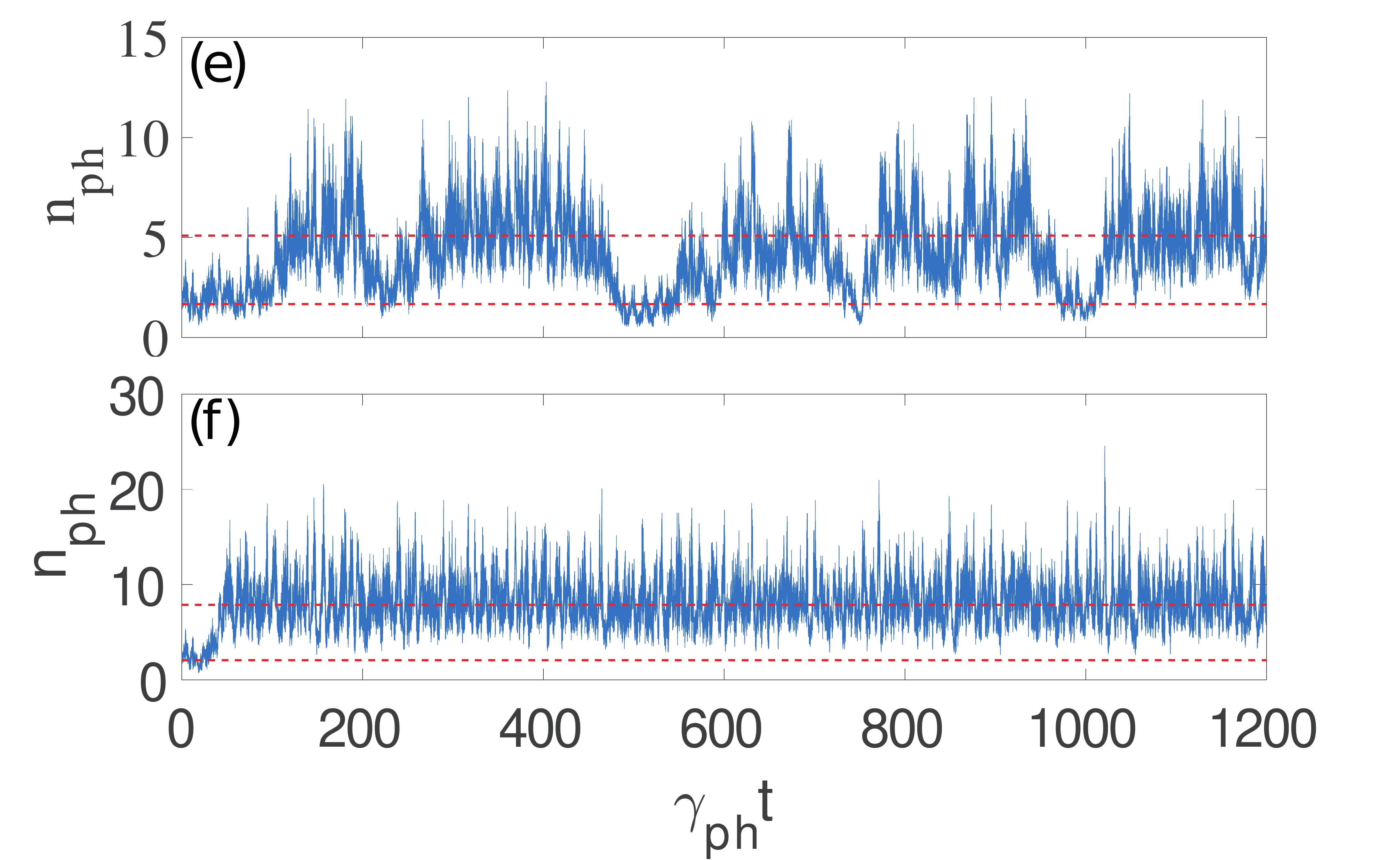}} 
\end{minipage}
\caption{~(a) The bistability map on the parameters plane of $\Delta $ and $\Omega $. Shaded region corresponds to the existence of the bistability. Color of the filling corresponds to the lower threshold of the driving field intensity within the bistability loop, see panel (b). (b) The photon population $n_{ph}$ and analytical quantum solution \eqref{Eq4_} for $\left\langle \phi ^{+} \phi \right\rangle $ for the parameters $\Delta =32\gamma_{ph}$, $\Omega =-49.2\gamma_{ph}$ (orange line), $\Omega =-49.1\gamma_{ph}$ (blue line), $\Omega =-49\gamma_{ph}$ (red line), $\Omega =-48.9\gamma_{ph}$ (green line) -- color points on the panel (c). (c) the ratio of the driving intensities corresponding to the bistability loop turning points. (d) The second order correlation function $g_{ph}^{(2)} $ dependence on the driving intensity $I_{d} $. The parameters are $\alpha =0.015\gamma_{ph}$, $\omega_{{\rm R}} =25\gamma_{ph}$, $\gamma_{{\rm ex}} =0.01{\rm \; ps}^{-1}$ and $\gamma_{{\rm ph}} =0.1{\rm \; ps}^{-1} .$ (e,f) A numerical simulations of the stochastic dynamics of the photon number predicted by the equations (A4). The initial state corresponds (e) to the red point 2 on the panel (b) and (f) to the red point 7.}
\label{Fig.figure1all02}
\end{figure}

Based on the these examples, we distinguish between two different dynamical regimes. In the first regime the dynamics is accompanied by the frequent jumps such as shown in Fig.~2e. In Fig.~2b this regime corresponds to the orange and blue curves. Since this behavior is caused by the quantum noise, this regime is referred to as a quantum case. In the bistability diagram, this regime arises close to the boundary of the bistability existence domain, see Fig.~2c, when the turning points of the $S$-shaped classical bistability curve are close to each other. The second case is characterized by the narrow region of the pump intensities where the quantum jumps occur. It corresponds to  the domain where the mean photon number \eqref{Eq4_} grows steeply. In this case, the region of the metastable solutions characterized by a single jump behavior illustrated by Fig.2f is typically much wider, see the red and green curves in Fig.2b. The second regime occurs far away from the boundary of the bistability existence domain (see Fig.~2c), where the turning points of the bistability loop move away from each other. Adopting the terminology from \cite{CRC} we call this case a quasiadiabatic regime implying quasi-equilibrium of the system.
In article \cite{CRC}, similar regime was found in the context of a dissipative phase transition when the number of particles in the system tends to infinity, i.e., in the thermodynamic limit with closing of the Liouvillian gap 
However, in our case, a similar regime occurs for small populations.

The differences between two regimes can be clearly illustrated with the use of the Wigner function representation shown in Fig.~3. The coexistence of two metastable states is indicated by a bimodal structure \cite{M} of the Wigner function illustrated in Fig.~3b and Fig.~3e. Note that when a single stable state exists, the Wigner function is localized and characterized by a single peak, see the panels (a), (c), (d) and (f) corresponding to the points 1, 3, 4 and 6 in Fig.~2b, respectively. For the quantum regime, in the particular case corresponding to the recurrent jumps between the points 2 and $2^\prime$ in Fig.~2b, one can see a pronounced overlap of the quasi-probability distributions corresponding to two bistable states, see Fig.3b. On the contrary, for the quasiadiabatic case, corresponding  to the points $5$ and $5^\prime$ in Fig.~2b the quasiprobability is localized in the two regions with almost no overlap between them, see Fig.~2e.

The quantum behavior of the photonic mode in the bistable regime is reflected in its statistics. In the region where the quantum average photon number grows steeply, we observe a peak of the second order correlation function $g_{ph}^{(2)}$, see Fig.~2d, which corresponds to effect of bunching  of photons emitted by the polariton system. Fig.~2d shows the second order correlation functions of photons for various laser detunings $\Omega$ corresponding to the solutions shown in Fig.~2b. We can see an increase of the peak amplitude accompanied by the decrease of its width as the laser intensity growth. In this case, the width of the peak of the second order correlation function is comparable with the width of the growth region of the quantum solution.

\begin{figure}
\begin{minipage}[h]{0.32\linewidth}
\center{\includegraphics[width=1\linewidth]{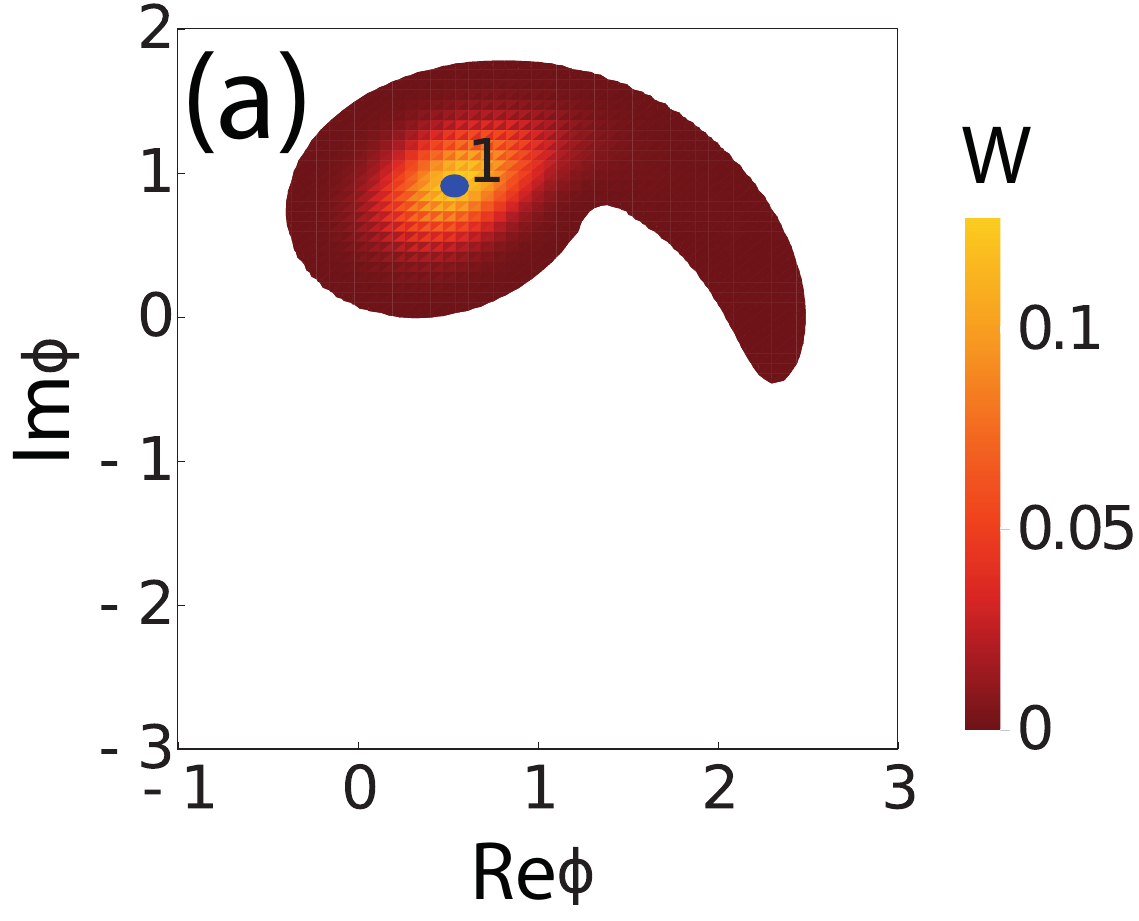}} 
\end{minipage}
\hfill
\begin{minipage}[h]{0.32\linewidth}
\center{\includegraphics[width=1\linewidth]{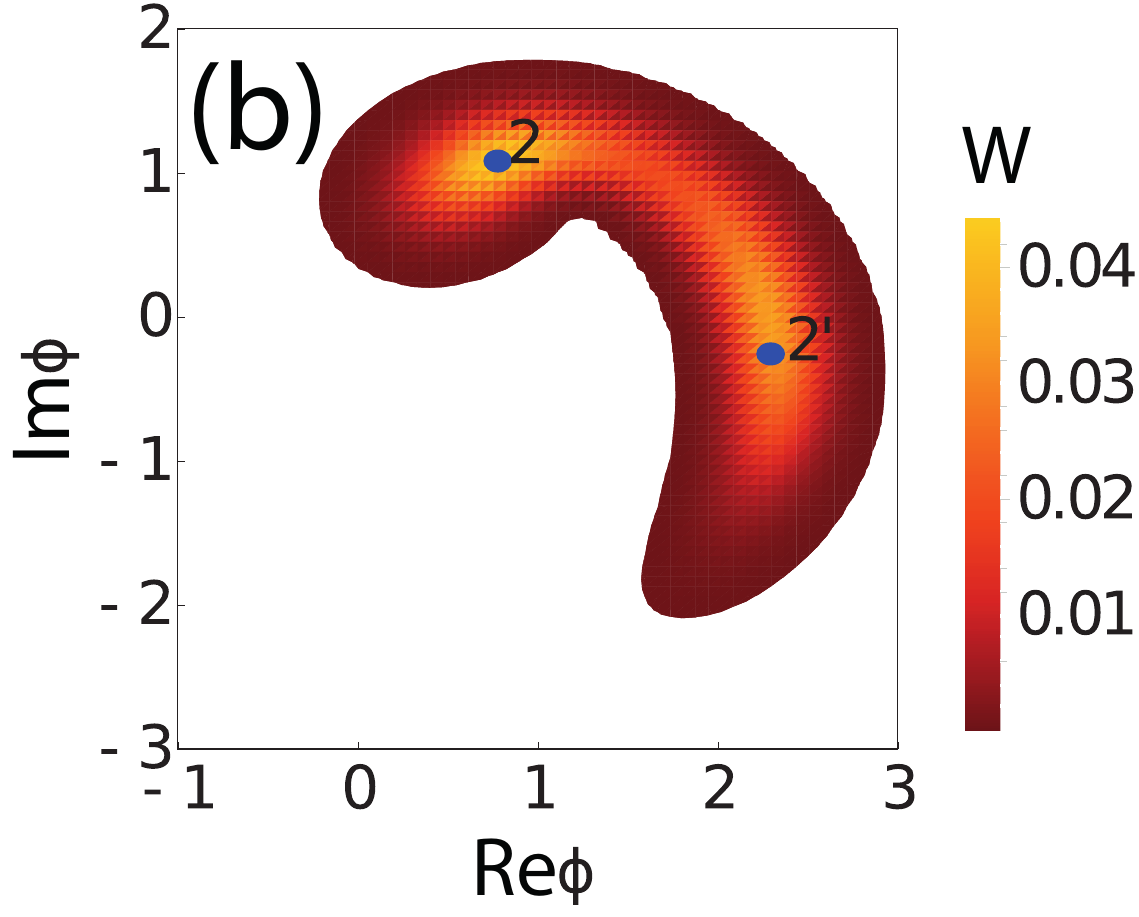}}
\end{minipage}
\hfill
\begin{minipage}[h]{0.32\linewidth}
\center{\includegraphics[width=1\linewidth]{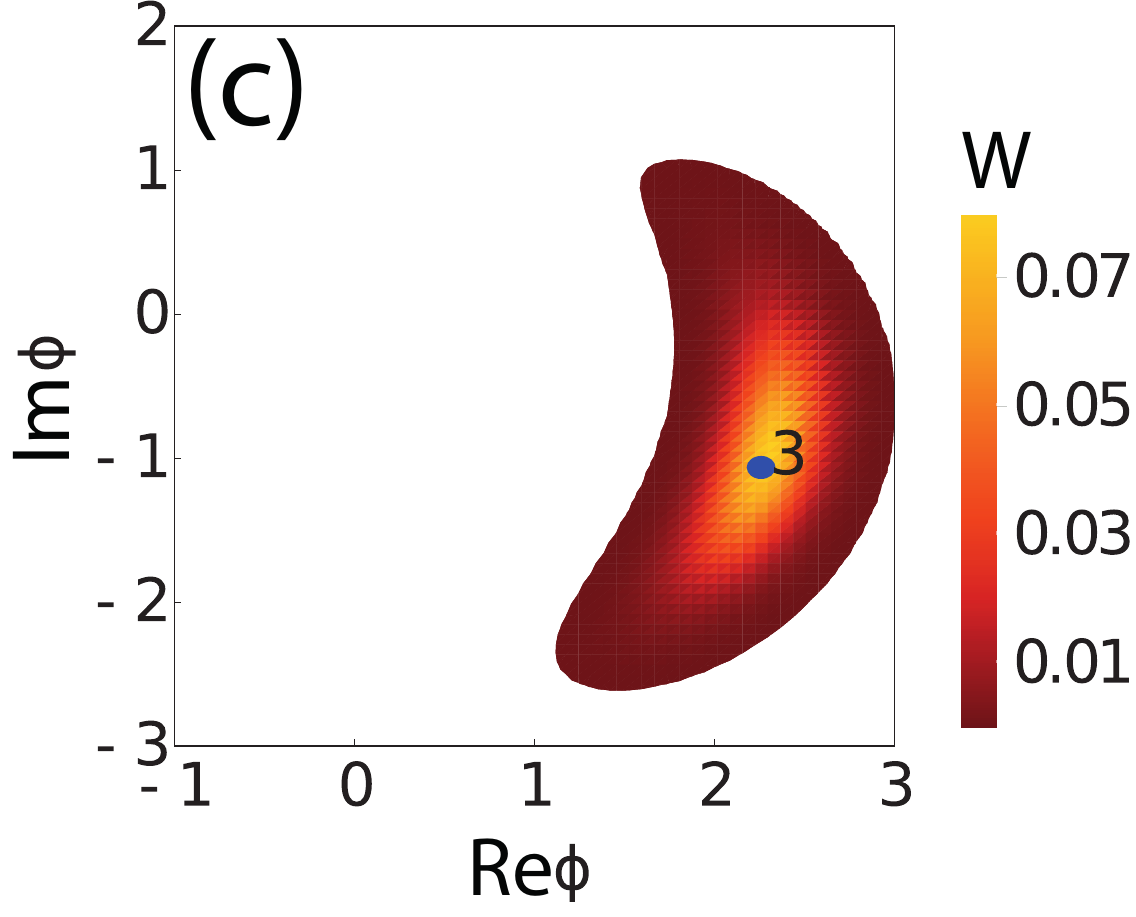}} 
\end{minipage}
\vfill
\begin{minipage}[h]{0.32\linewidth}
\center{\includegraphics[width=1\linewidth]{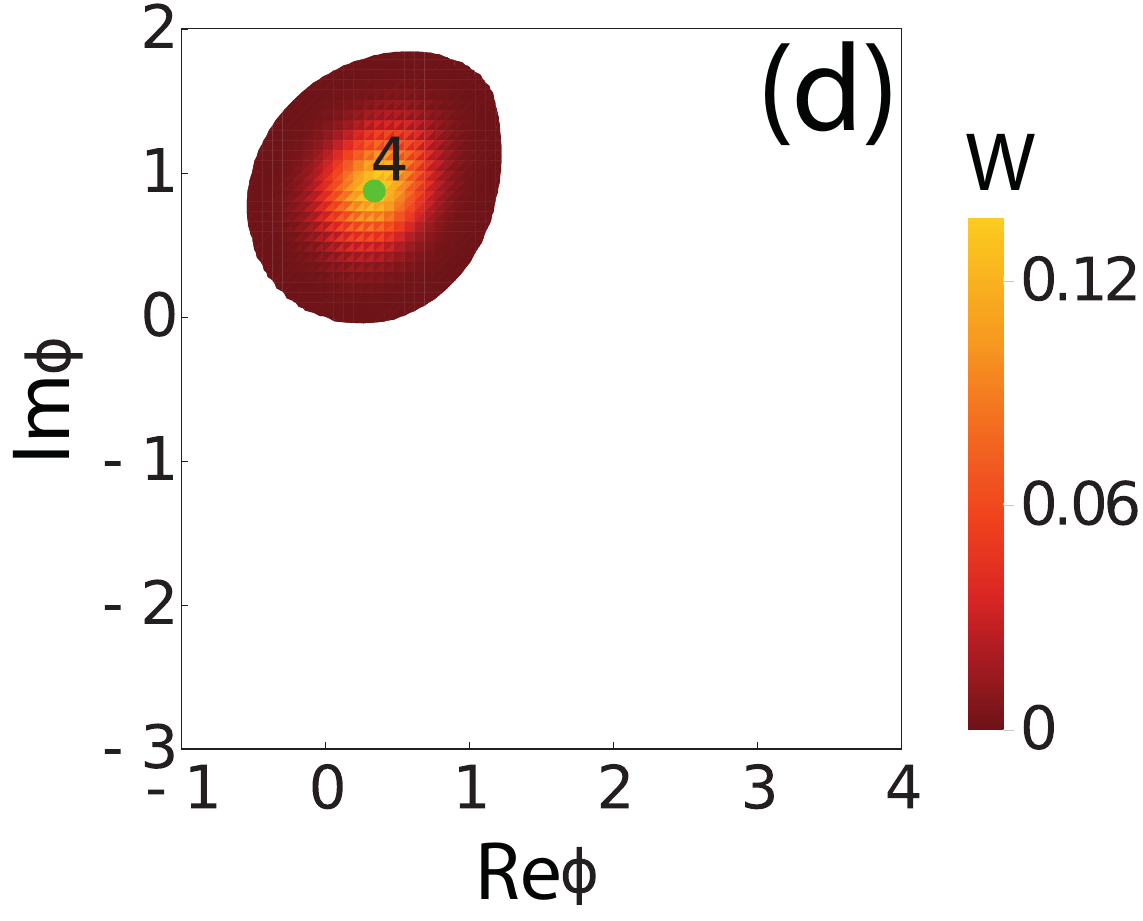}} 
\end{minipage}
\hfill
\begin{minipage}[h]{0.32\linewidth}
\center{\includegraphics[width=1\linewidth]{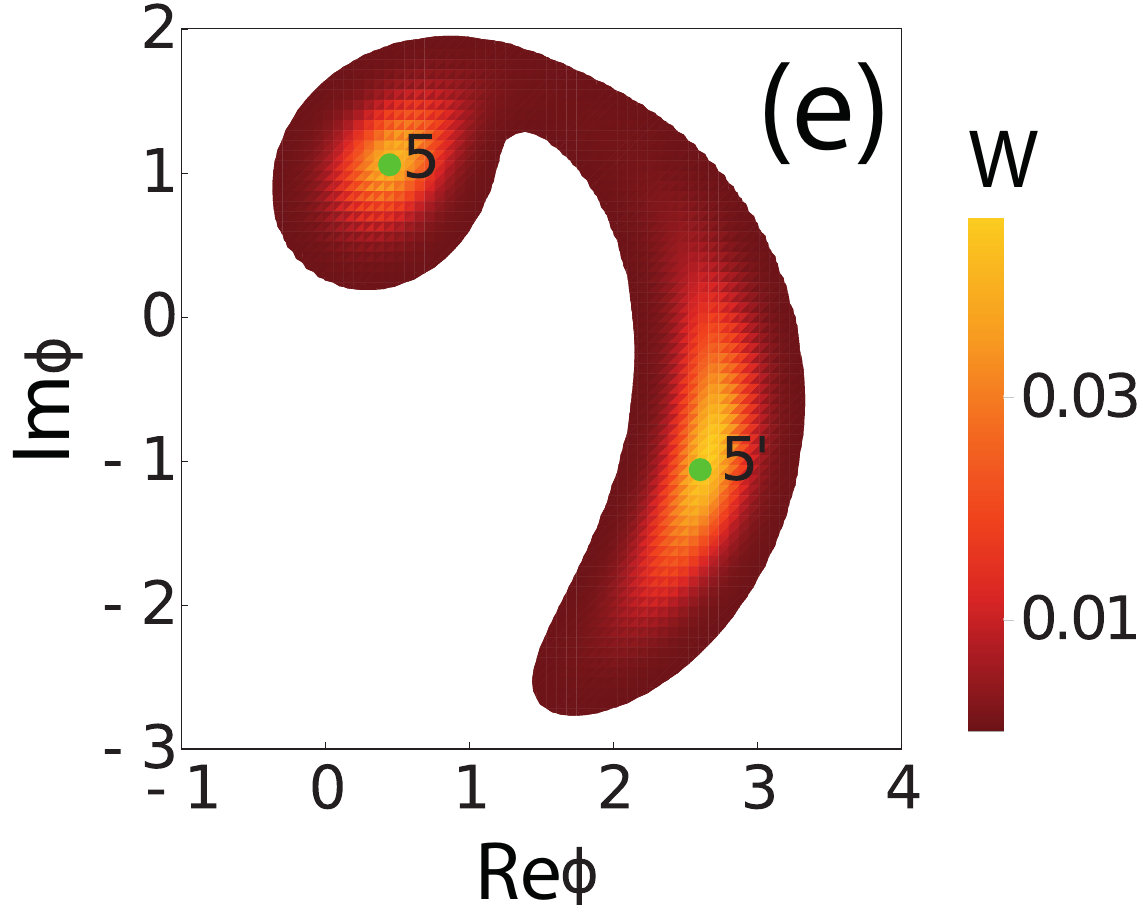}} 
\end{minipage}
\hfill
\begin{minipage}[h]{0.32\linewidth}
\center{\includegraphics[width=1\linewidth]{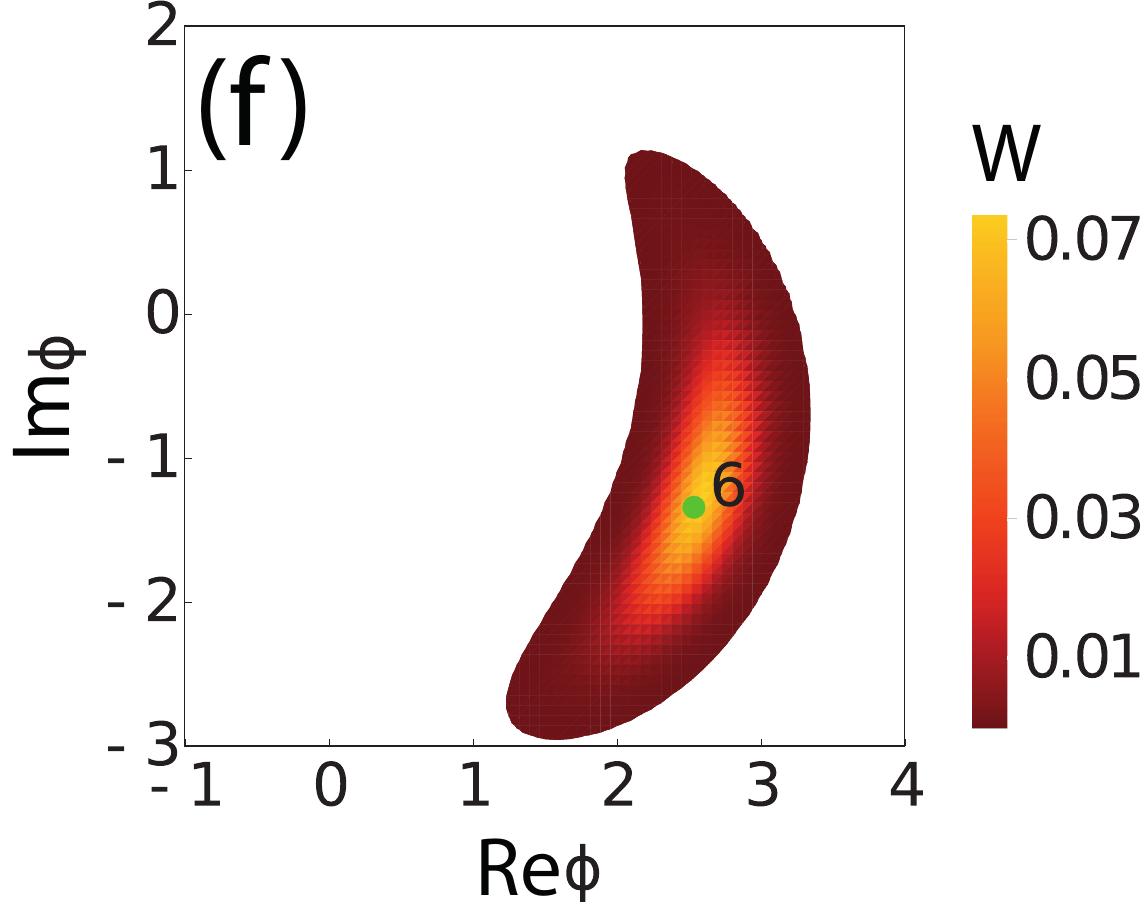}} 
\end{minipage}
\caption{~The Wigner function distribution $W({\rm Re}\phi,{\rm Im}\phi)$ characterizing the photonic fraction. Panels (a), (b) and (c) correspond to different positions on the blue curve of Fig.~2b indicated by the points $1$, $2$ and $3$ respectively. The detunings are $\Omega = -49.1\gamma_{ph}$ and $\Delta = -32\gamma_{ph}$.  Panels (d), (e) and (f) correspond to the points 4, 5 and 6 on the green curve respectively. The detunings are $\Omega = -48.9\gamma_{ph}$, $\Delta = -32\gamma_{ph}$.}
\label{Fig. The Wigner function distribution}
\end{figure}

Figure 4 shows a map of the second order correlation function on the $\Omega $ and $\Delta $ parameter space for various driving intensities: (b) $I_{d}/\gamma_{ph}^{2}=70$, (a),(c) $I_{d}/\gamma_{ph}^{2}=100$, and (d) $I_{d}/\gamma_{ph}^{2}=200$. Panels (b-d) show the region in the vicinity of the resonance of the lower polariton (LP) branch (the green dotted curves), while the panel (a) is focused on the upper polariton (UP) branch. One can see a bright narrow band which corresponds to the peaks of the $g^{(2)}_{\rm ph}$ function while the background value is $g^{(2)}_{\rm ph}=1$.  It follows the shape of the LP branch resonance curve. Note that the a narrow band of bunching states as well as the peak magnitude shifts towards larger driving detunings $\Omega$ as the pump intensity increases. Besides, the peak of the second order correlation function is more pronounced for a positive detuning, $\Delta>0$. This is because the quantum fluctuations of the lower polaritons play a dominant role near the LP resonance (in contrast to the fluctuations of the upper polaritons which are suppressed in this domain). The fluctuations are more pronounced for positive detunings since in this region the exciton fraction  of the lower polariton is dominating over the photon fraction. It is reflected in the value of the Hopfield coefficients $C_{\rm ex}=\sqrt{\frac{1}{2}\left(1-\frac{\Delta}{\sqrt{\Delta^{2}+\Omega_{R}^{2}}}\right)}$ and $C_{\rm ph}=\sqrt{\frac{1}{2}\left(1+\frac{\Delta}{\sqrt{\Delta^{2}+\Omega_{R}^{2}}}\right)}$ which determine the exciton and photon fractions in the polariton state, respectively, and depend on the detuning $\Delta$ and the Rabi splitting $\omega_{R}$. In particular, for the positive detuning $C_{\rm ex} >C_{\rm ph}$.

\begin{figure}
\begin{minipage}[h]{0.49\linewidth}
\center{\includegraphics[width=1\linewidth]{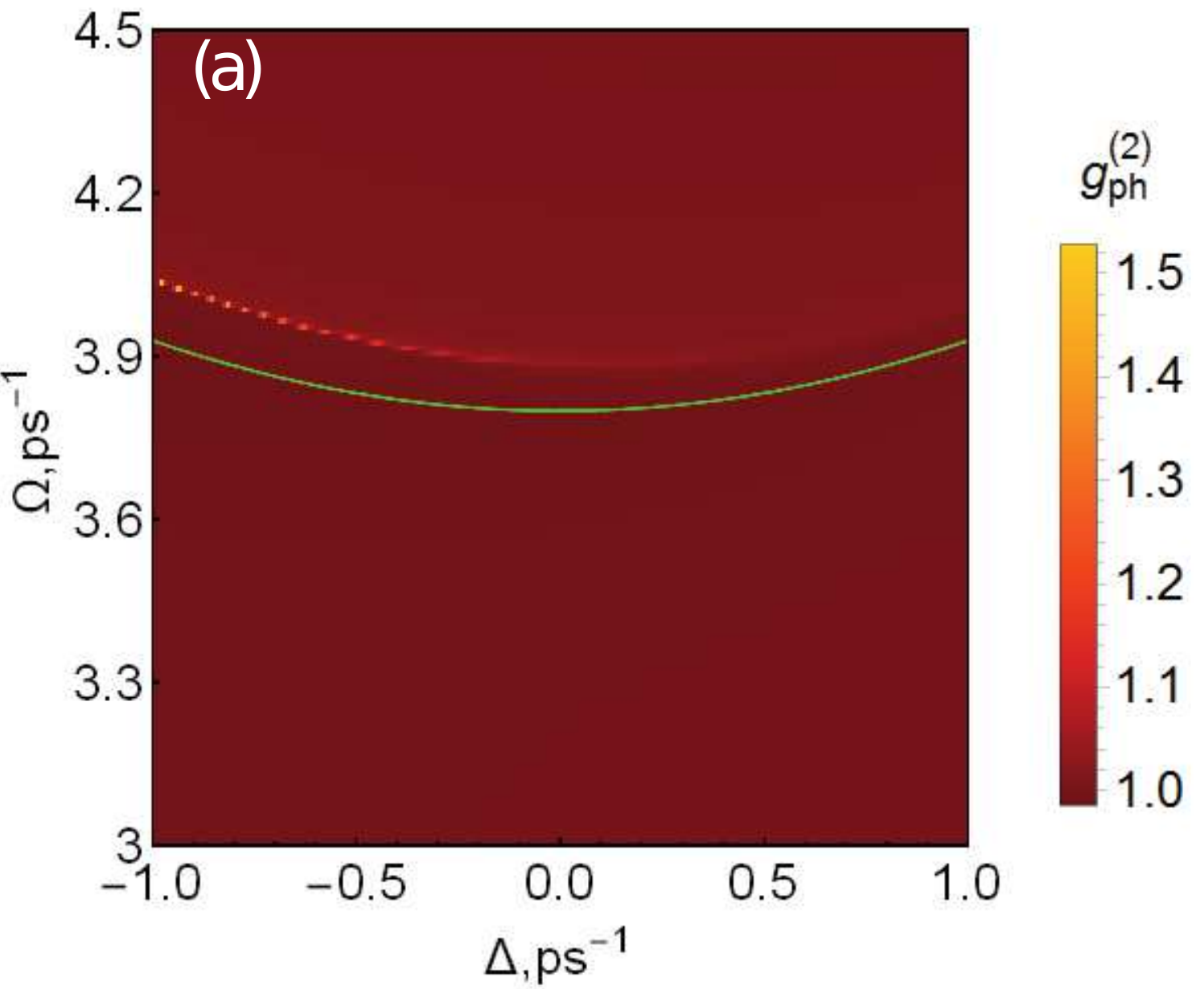}} 
\end{minipage}
\hfill
\begin{minipage}[h]{0.49\linewidth}
\center{\includegraphics[width=1\linewidth]{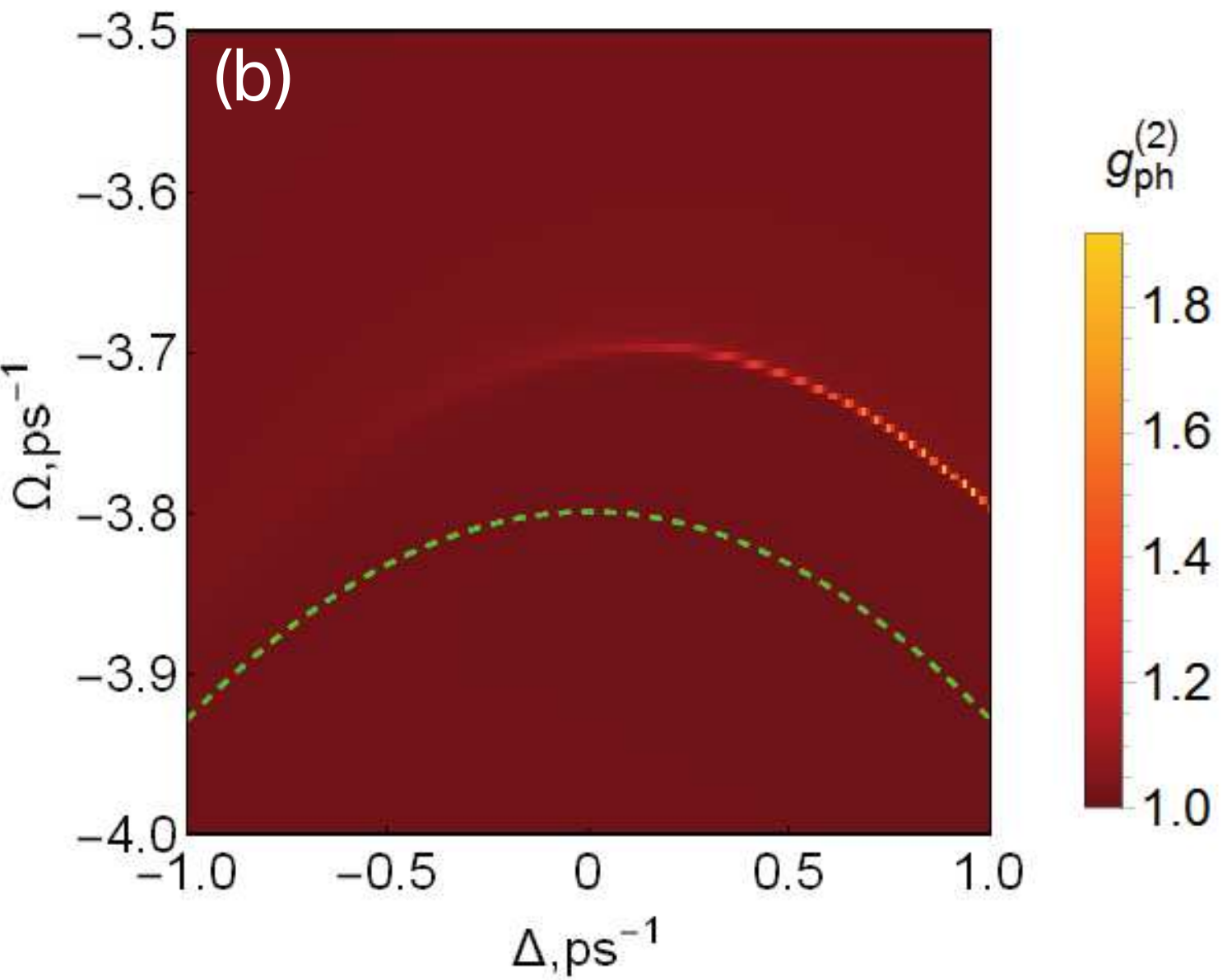}}
\end{minipage}
\vfill
\begin{minipage}[h]{0.49\linewidth}
\center{\includegraphics[width=1\linewidth]{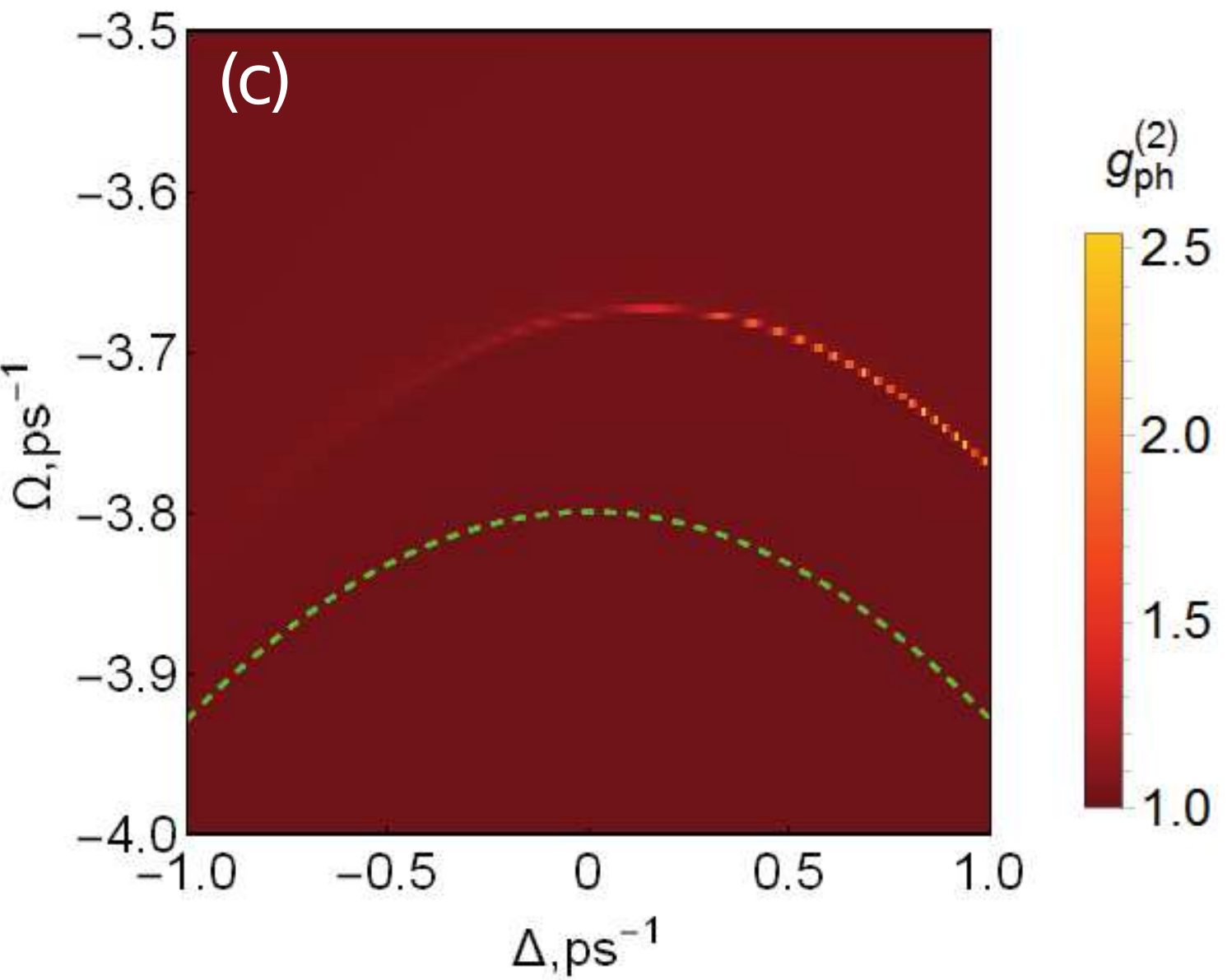}} 
\end{minipage}
\hfill
\begin{minipage}[h]{0.49\linewidth}
\center{\includegraphics[width=1\linewidth]{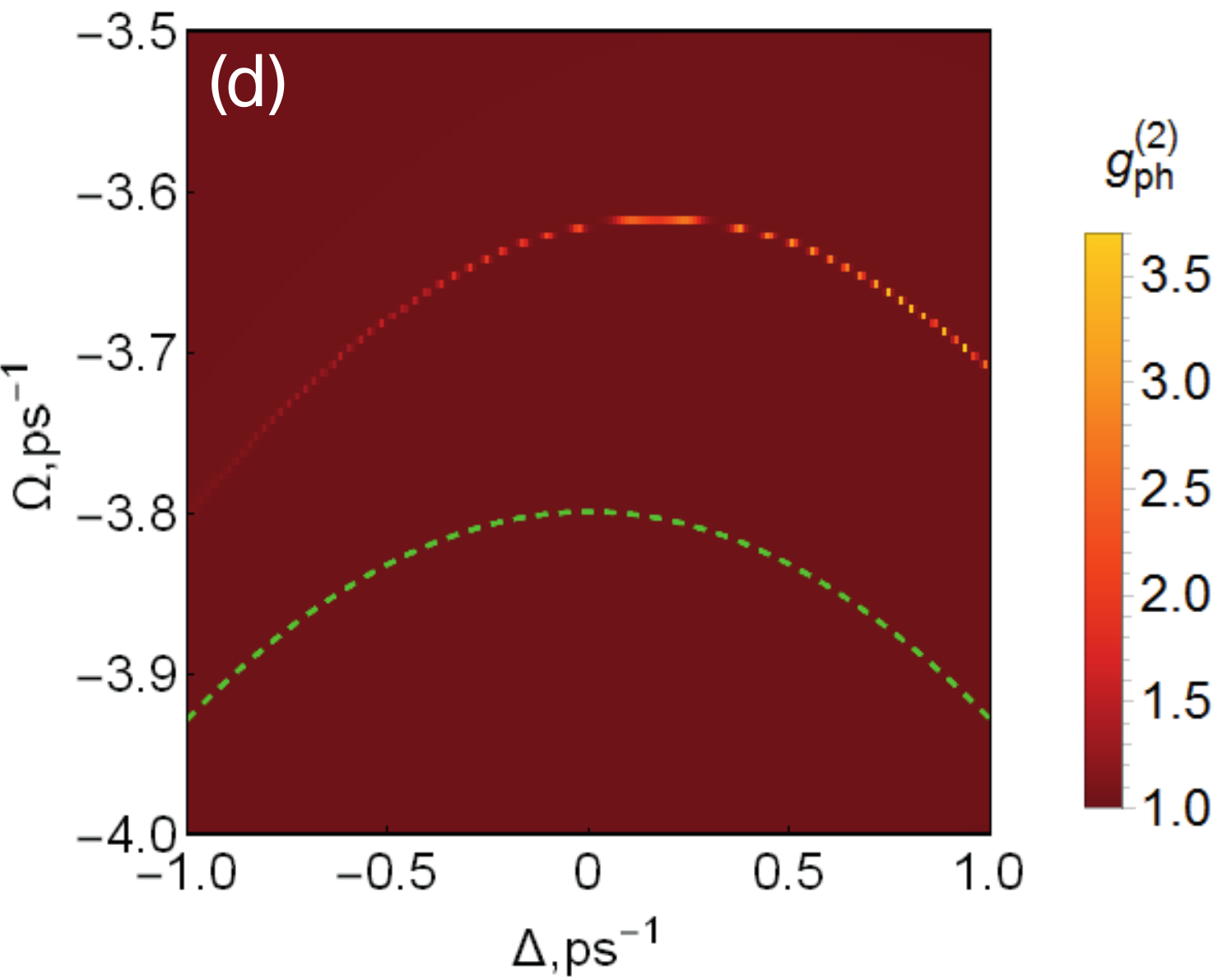}} 
\end{minipage}
\caption{The maps of the  second order correlation function of photons in the vicinity of the upper polariton branch (a) and of the lower polariton branch (b)-(d). The polariton frequencies are indicated with green dotted lines. The different driving intensity is (a) $\sqrt{I_{d}}/\gamma_{ph}=100$, (b) $I_{d}/\gamma_{ph}^{2}=70$, $I_{d}/\gamma_{ph}^{2}=100$ and (d) $I_{d}/\gamma_{ph}^{2}=200$.}
\label{Fig. The second order correlation function}
\end{figure}

Summarizing this section, we emphasize a crucial role of the quantum noise in the optical bistability phenomenon in semiconductor microcavities. The noise-induced quantum jumps between metastable states wash out the hysteresis loop and lead to the non-classical statistics of the emitted light. However in practice, the stochastic behavior of optical photons is detectable only close the bistability threshold, i.e. when the classical hysteresis loop is narrow. For the wide loops, the time spent by the system in the metastable state quickly grows with the increase of the loop width and can become impractically long as it was discussed in \cite{CRC}. In contrast to the previous studies, our approach demonstrates that the non-classical  statistics of the emitted light can be detected for a wide range of the laser driving detunings in the vicinity of both lower and upper polariton resonances. However, the range where the noise-induced behavior takes place is not limited solely to the domain of bistability. In the following section we demonstrate the phenomena of photon bunching and giant antibunching which occur far from the bistability existence domain.


\section{Triple resonance region}

In this section we focus on the parameter region close to a triple resonance between the exciton, photon and the pump laser frequencies, $\Omega=0$ and $\Delta=0$. In this case we observed the bunching, giant bunching and the antibunching  phenomena. The $\left(\Omega,\Delta\right)$ -- map of the second order correlation function  $g_{{\rm ph}}^{\left(2\right)} (0)$ in the vicinity of the triple resonance is shown in Fig.~5a. One can see a pronounced light domain indicating the effect of giant bunching close to the line $\Omega\approx-\Delta$. This behavior should be attributed to the imbalance in the photon and exciton populations. In particular, in the region of the giant bunching, the exciton population dominates over the photon population $n_{ex}/{n_{ph}} \gg 1 $ as it is shown in Fig.~5b. Therefore, small quantum fluctuations of the exciton mode cause large quantum fluctuations of the photon mode according to the slaving principle \cite{H2013}, see Eq.~\eqref{Eq3_}. This is illustrated in Fig.~5b demonstrating the map of the $n_{ex}/{n_{ph}}$ ratio. The maximum in Fig.~5b follows the region of the giant photon bunching effect  $g_{{\rm ph}}^{\left(2\right)} (0) \gg 1$ observed in Fig.~5a. Note, that for the considered parameters the value $g_{{\rm ph}}^{\left(2\right)} (0)$ in the giant bunching regime reaches up to $10^5$ as it is illustrated in Fig. 5c demonstrating a cross-sections of the map Fig.~5a for fixed values of the exciton-photon detuning.

\begin{figure}
\begin{minipage}[h]{0.49\linewidth}
\center{\includegraphics[width=1\linewidth]{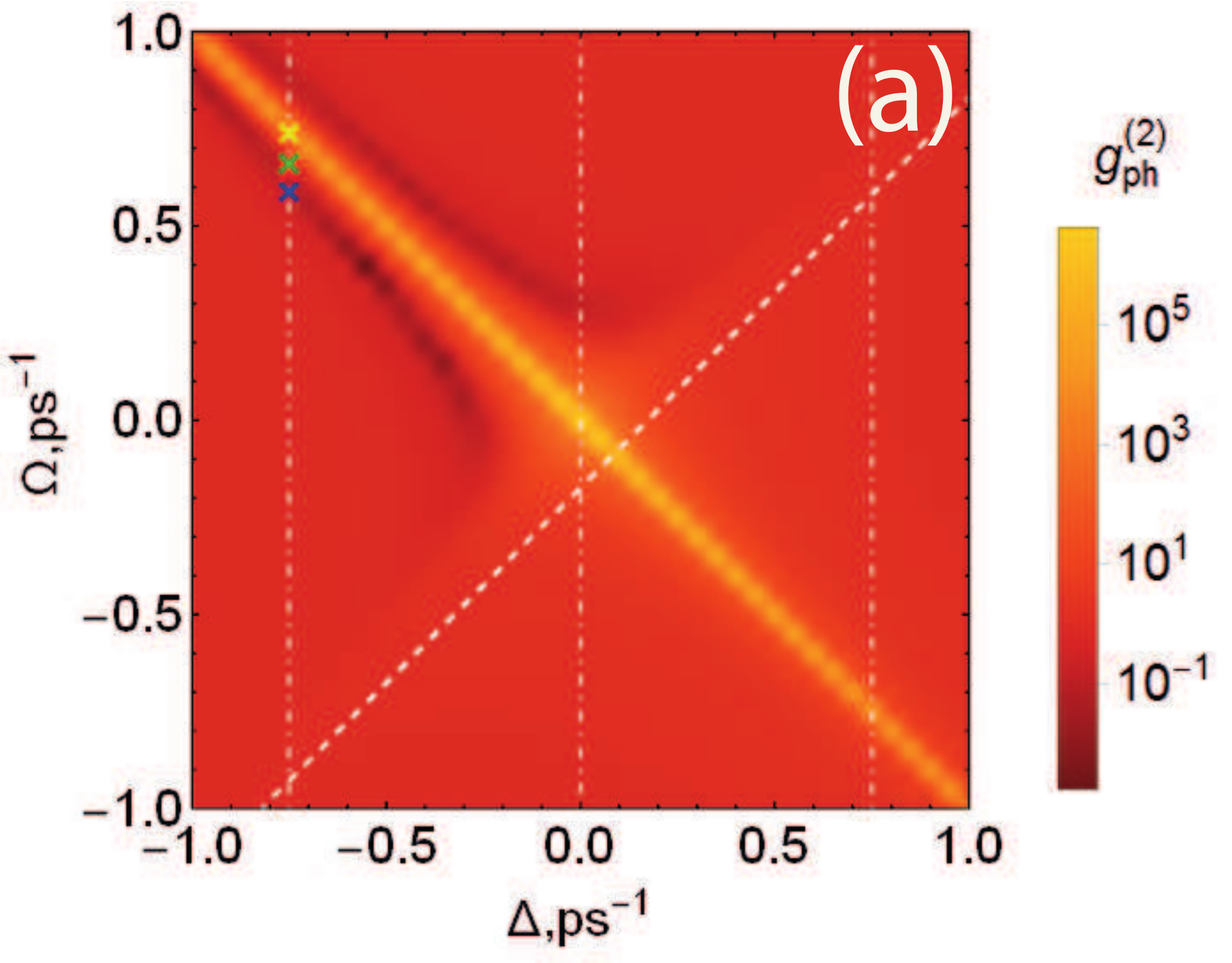}} 
\end{minipage}
\hfill
\begin{minipage}[h]{0.49\linewidth}
\center{\includegraphics[width=1\linewidth]{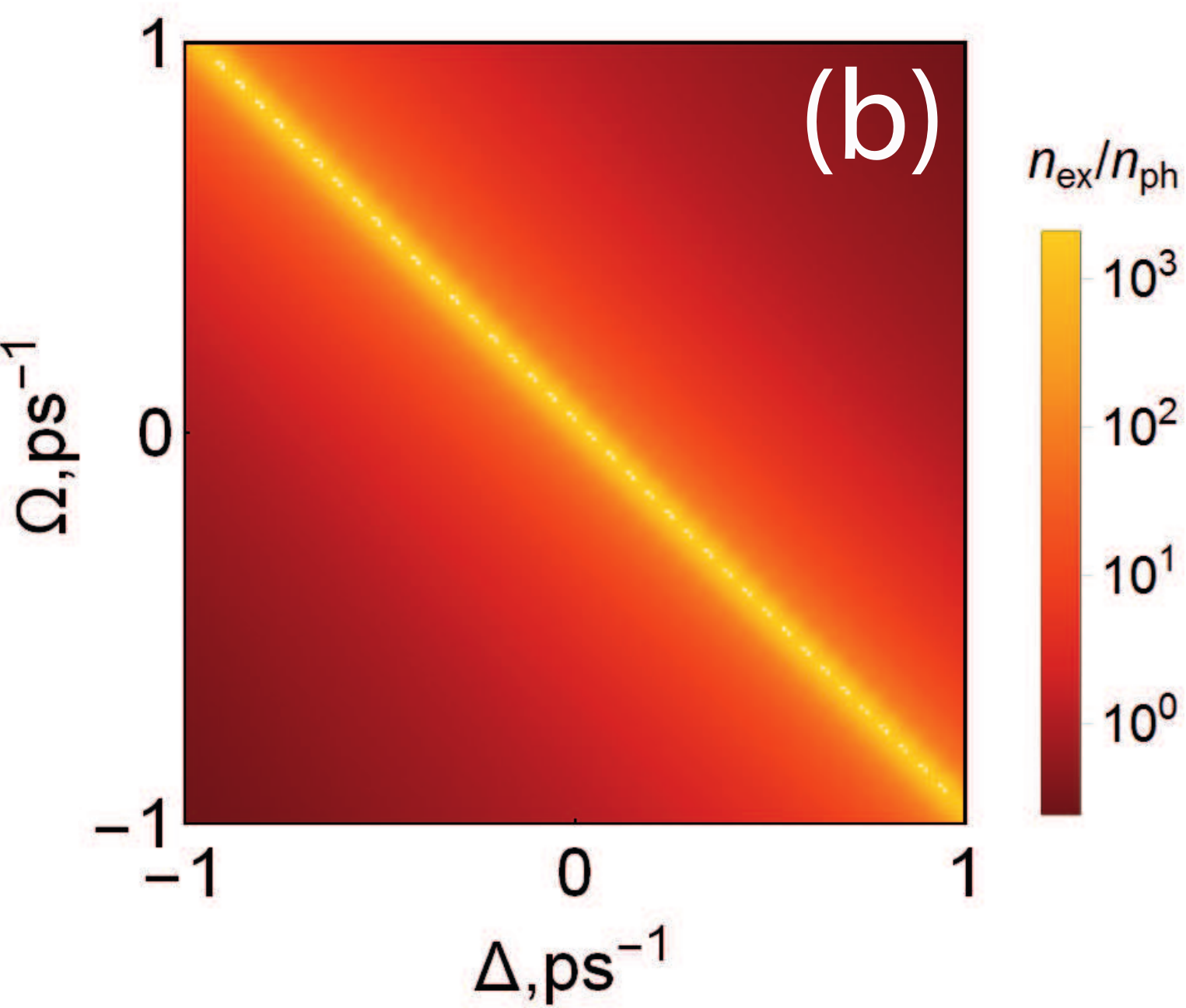}}
\end{minipage}
\vfill
\begin{minipage}[h]{0.49\linewidth}
\center{\includegraphics[width=1\linewidth]{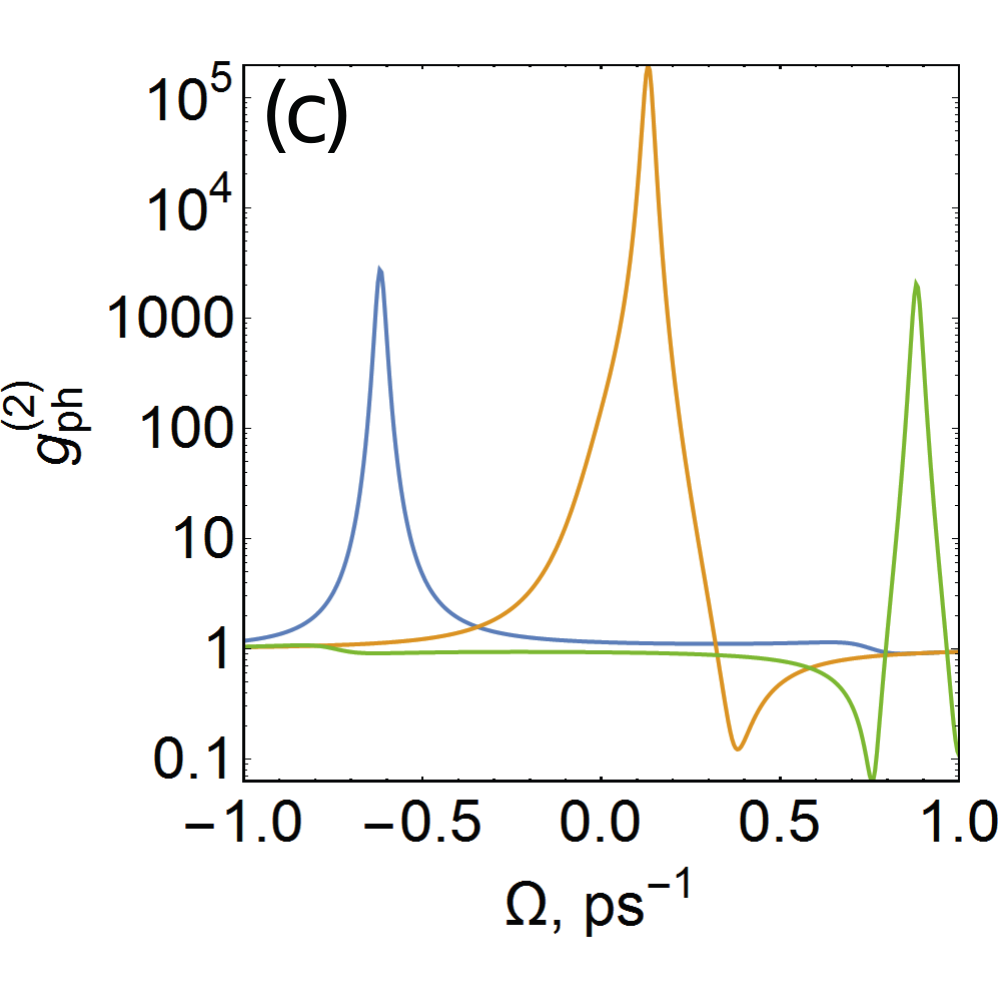}} 
\end{minipage}
\hfill
\begin{minipage}[h]{0.49\linewidth}
\center{\includegraphics[width=1\linewidth]{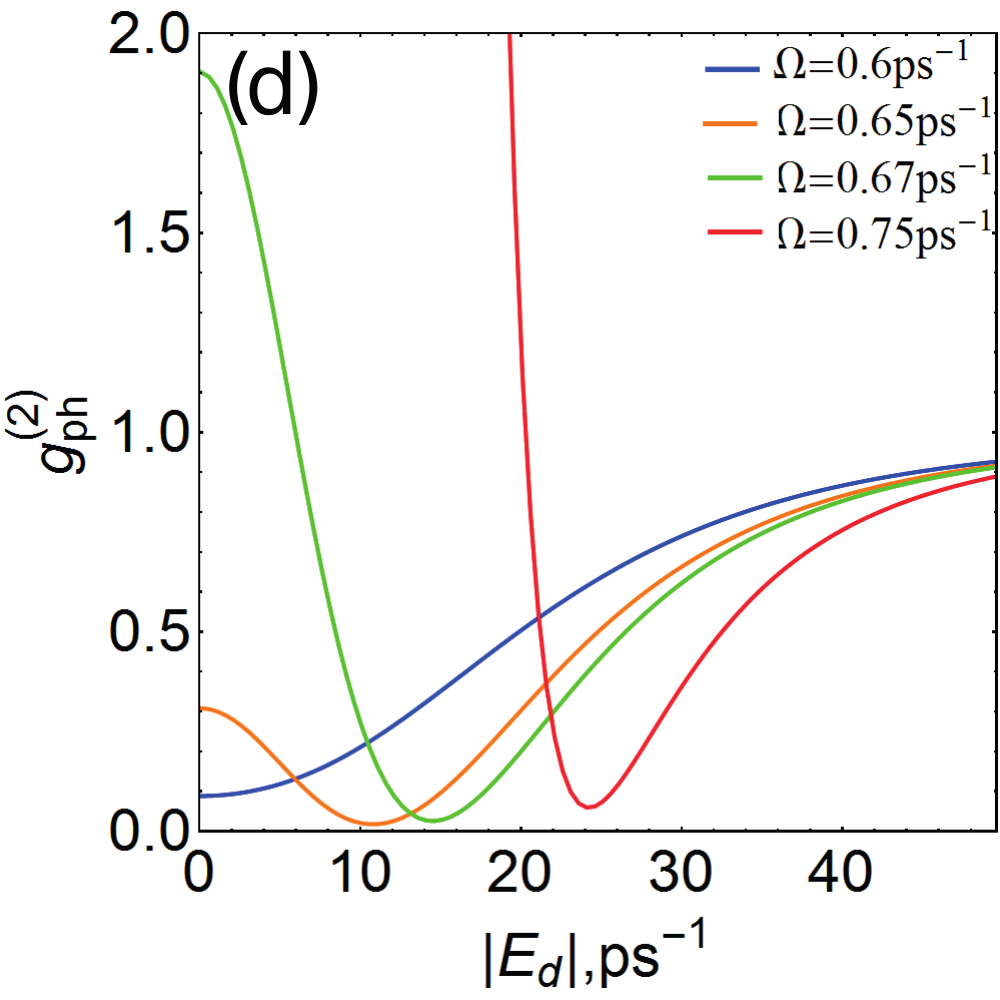}} 
\end{minipage}
\caption{~(a) The zero-delay second order correlation function of photons  $g_{{\rm ph}}^{\left(2\right)} (0)$ for various values of the driving detuning $\Omega $  and the exciton-photon detuning $\Delta $. (b) The ratio between populations of the excitonic and the photonic modes $n_{ex}/{n_{ph}}$ for the same parameter plane as in the panel (a); (c) The dependence of $g_{{\rm ph}}^{\left(2\right)} (0)$ on the driving detuning $\Omega $ for the fixed values of the exciton-photon detuning, shown by the vertical dot-dashed lines in the panel (a), $\Delta =-7.5\gamma_{ph}$ (green line), $\Delta =0$ (orange line), and $\Delta =7.5\gamma_{ph}$ (blue line). (d) The dependence of $g_{{\rm ph}}^{\left(2\right)} (0)$ on the driving amplitude $|E_{{\rm d}}|$ for $\Delta =-7.5\gamma_{ph}$, and the fixed values of the laser detuning (colour crosses in the panel (a)), $\Omega = 6\gamma_{ph}$ (blue line), $\Omega = 6.5\gamma_{ph}$ (orange line), $\Omega = 6.7\gamma_{ph}$ (green line), and $\Omega = 7.5\gamma_{ph}$ (red line).}
\label{Fig. The giant bunching and antibunching effects}
\end{figure}

For negative detunings, the effect of giant bunching adjoins the region with $g_{{\rm ph}}^{\left(2\right)} < 1$, where the cavity photons are in the antibunching regime. The transition from the bunching to the antibunching statistics with the variation of the driving laser frequency is clearly seen in Fig.~5c. A driving field intensity-dependence of $g_{{\rm ph}}^{\left(2\right)}$  is shown in  Fig.~5d for different exciton-photon detunings: $\Omega = 6\gamma_{ph}$, $\Omega =6.5\gamma_{ph}$, $\Omega =6.7\gamma_{ph}$, and $\Omega = 7.5\gamma_{ph}$. the antibunching is typically observed for the moderate driving strength while in the limit of strong driving the cavity photon field approaches the coherent statistics, $g_{{\rm ph}}^{\left(2\right)}\approx 1$.

\section{Non-classic states]}

Previously, the effect of antibunching was studied for polaritons formed in a quantum box in the context of so-called polariton blockade \cite{VCC}. In particular, the antibunching phenomenon was predicted to occur only at sufficiently large nonlinear interaction strength which requires a microcavity with a very small mode volume \cite{VCC}. However, reducing the mode volume down to a subwavelength size has a drawback of the degradation of the microcavity quality factor which corresponds to the increase of the cavity mode dissipation rate.
Here, we demonstrate that the antibunching effect can be obtained with the small values of the nonlinear interaction strength which are achievable in the state of the art pillars microcavities. This is demonstrated in Fig.~6a which shows the second order correlation function of photons for different values of the nonlinear interaction strength normalized to $\alpha_{0} =0.015\gamma_{ph}$ which correspond to the value used for the rest of the calculations in the paper. The photon antibunching effect can be observed even at weak nonlinearities.

\begin{figure}
\begin{minipage}[h]{0.49\linewidth}
\center{\includegraphics[width=1\linewidth]{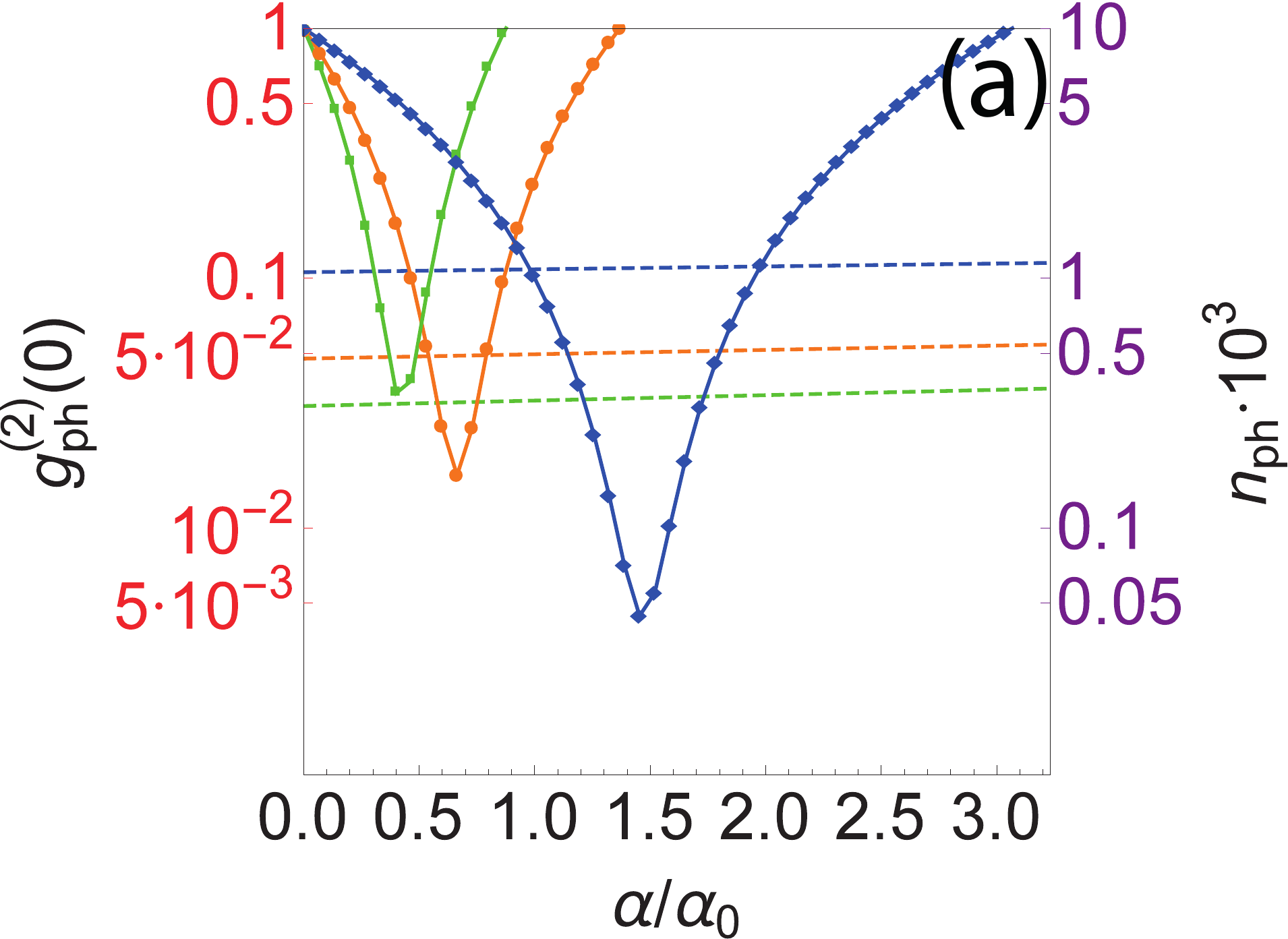}}
\end{minipage}
\hfill
\begin{minipage}[h]{0.48\linewidth}
\center{\includegraphics[width=1\linewidth]{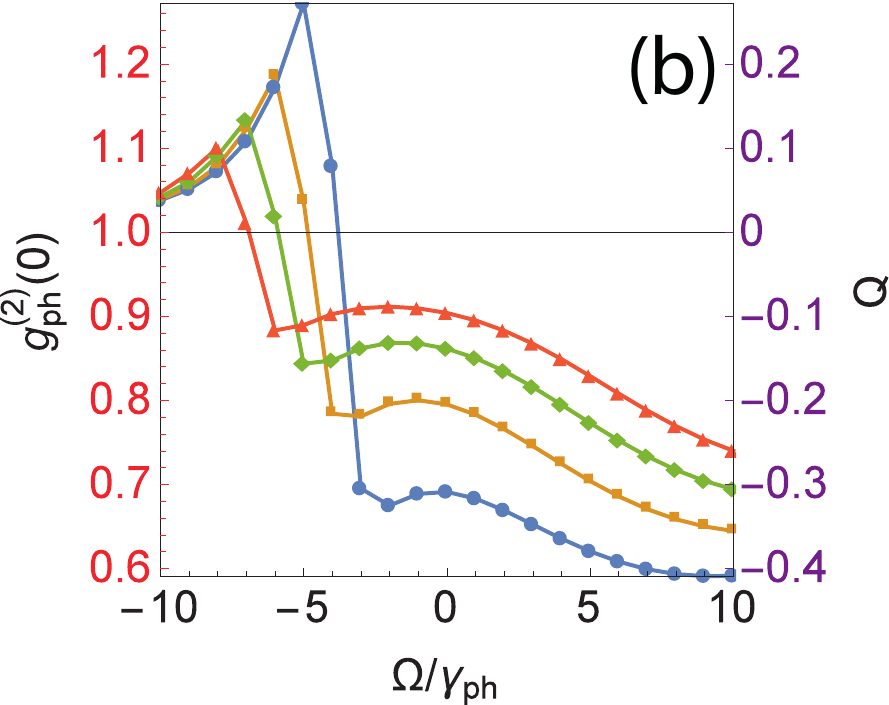}}
\end{minipage}
\caption{~(a) The second order correlation function of the photons versus the nonlinearity parameter for the following parameters $\Delta =-7.5\gamma_{ph}$, $\Omega=6\gamma_{ph}$ -- blue line, $\Omega=6.5\gamma_{ph}$ -- the orange line, $\Omega=6.7\gamma_{ph}$ -- the green line.~(b) The second order correlation function of the photons  and the Mandel parameter versus the laser detuning for average value of photons $\left\langle \phi ^{+} \phi \right\rangle = 1$ for the following parameters $\Delta =-4\gamma_{ph}$ -- blue line, $\Delta =-5\gamma_{ph}$ -- orange line, $\Delta =-6\gamma_{ph}$ -- green line, $\Delta =-7\gamma_{ph}$ -- red line. 
}
\label{Fig. The antibunching}
\end{figure}

The strongly pronounced antibunching effect can serve as a manifestation of the effect of quantum blockade. This effect can be used to create single photon sources. That is why the relevant phenomenon of the polariton blockade is of a great interest now \cite{delteil2019}.
Another important question is whether the photon field emitted by the microcavity reflects the statistic of the intra-cavity polaritons. To answer this question, we address the statistical properties of the second constituent of polaritons –- the quantum well excitons. The effects of antibunching can be also observed for the exciton mode. However, in contrast to the case of photons, the exciton antibunching is weak due to the large number of excitons. In fact, the $g^{(2)}_{\rm ex}$ map follows that of the photon field though the peak and the deep amplitude values of the correlation function are much less. For the antibunching region $g^{(2)}_{ex}$ is typically a little less than one. 
The quantum statistics of the polaritons of the lower branch combines the statistics of photons and excitons, and due to the domination of the exciton field, is close to the exciton statistics. Namely, one can easily check that the $g^{(2)}_{LP}\geq(g^{(2)}_{ph}+g^{(2)}_{ex})/2$ is true in the region close to triple resonance. It means that polaritons do not exhibit a noticeable antibunching effect in contrast to photon. Therefore, it is necessary to emphasis that the quantum statistics of polaritons does not have to coincide with the statistics of the microcavity radiation. When a polariton is emitted by a microcavity, polariton state collapses to a photonic state inheriting the energy and momentum from the polariton. However, the statistics of the emitted photon field can be completely different.
%
%
%

The antibunching effect occurs when quantum anticorrelation of photon pairs, which leads to a sub-Poissonian distribution of the number of photons. However, the antibunching of photons is generally not completely equivalent to the notion of sub-Poissonian statistics. For example, non-classical statistics can be found even when $g_{ph}^{2}(0)>1$ (it which manifests as violation of inequality $g_{ph}^{2}(\tau)>g_{ph}^{2}(0)$). The characteristic of the deviation of radiation statistics from Poisson can be the Mandel parameter \cite{Mandel}:
\begin{equation} \label{Eq6_}
Q=\frac{\left\langle \Delta n^{2} \right\rangle}{\left\langle n \right\rangle}-1,
\end{equation}
we can rewrite the Mandel parameter to the second order $g_{ph}^{2}(0)$ as follows
\begin{equation} \label{Eq7_}
Q=\left\langle n_{ph} \right\rangle (g_{ph}^{(2)}(0)-1).
\end{equation}

On the Fig.~6a we can see the antibunching effect but at the same time we have small average values of the photon number which show dashed lines on the right axes in Fig.~6a. In these cases we observation sub-Poissonian statistic of photons $g_{ph}^{2}(0)<1$ but close to Poisson statistic since the Mandel $Q$ close to zero.
We can fix the average number of photons $\left\langle n_{ph}\right\rangle =1$ by choosing the appropriate pump intensity. These cases with  $g_{ph}^{2}(0)$ shows in Fig.~6b. We can see that for $\Omega>0$ the antibunching effect with statistic of photons close to the sub-Poissonian distribution since $Q<0$. On the right axis in the Fig.~6b shows the Mandel parameter which are demonstrated pronounced the sub-Poissonian statistic within $Q<0$.

\section{Conclusion}

We studied in detail the quantum behavior of exciton polaritons formed in the micropillar cavity. We obtained a quantum solution for an exciton polariton system using the quantum phase space methods. Namely, the quantum statistical averages for $g_{{\rm ph}}^{\left(2\right)}$ was obtained. The effect of quantum fluctuations on bistability is analyzed. We found quantum jumps between the states corresponding to the upper and lower branches of the bistability curve. Previously, these stochastic jumps between bistable states were observed experimentally \cite{KFA,AST,WEL}. We distinguish between several regimes in the bistable region: 1) a quantum regime with nontrivial quantum behavior and smooth boundaries between bistable states and 2) the quasiadiabatic case with the presence of metastable states and sharp boundaries between them. The peak of the second order correlation function corresponding to the quantum statistics of the micropillar radiation was observed. It should be noted that the approaches developed in \cite{RCS,FSH,CS} used a single-mode approximation that takes into account only the lower polariton branch, when the pump frequency is tuned close to the resonance with the frequency of the lower polariton branch, and the contribution of the upper polariton the branch is neglected. Therefore, our work expands the scope of system parameters where the effect of the quantum noise on the statistical properties of the exciton-photon system can be analyzed.

In the region of the triple resonance, we discovered the  non-classical behavior of the photons, namely, the antibunching and the giant bunching phenomena. Moreover, the antibunching effect is observed even at small values of the non-linearity parameter. These results can be used in quantum technologies. For example, for creation of a polariton logic elements (example, polariton trigger proposed in article \cite{Quantpoltrig}) or qubits operating under bistablity conditions.
The recent experimental studies \cite{munoz2019} demonstrated that for the realization of a strong antibunching with semiconductor polaritons, the relatively strong interparticle interactions are required. In our case, this effect arises due to the fact that small quantum noise of the exciton mode, according to the slaving principle, \eqref{Eq3_}, induces strongly pronounced quantum effects in the photon mode. That is why the effects of strong antibunching appear at very low values of nonlinear parameter.

%
\medskip

The Authors acknowledge I.~Yu.~Chestnov for the fruitful discussions and the review of the manuscript. The research was supported by the Ministry of Science and Higher Education of the Russian Federation under Agreement No. 075-15-2019-1838. The work was also supported by the RFBR within the framework of the scientific project No. 20-02-00515.

\section*{Author contribution statement}
T. Khudaiberganov has carried out the analytical and numerical calculations, and wrote this paper and obtained physical results. All the authors contributed to the analysis and interpretation of the obtained results as well as to the final form of the manuscript.

\onecolumngrid
\appendix*
\section{Appendix A:~Derivation of basic equations based on the P-representation method}

Using the general $P$-representation and passing from operators to the $c$-numbers we turn from master equation \eqref{Eq2_} to the Fokker-Planck equation for the $P$-function \cite{DGG}:

\setcounter{equation}{0}

\begin{equation*} \label{EqA1_}
\begin{array}{l}{\frac{\partial P}{\partial t} =[-\frac{\partial }{\partial \phi } \left(-\left(i\Delta _{ph} +\gamma _{ph} \right)\phi +\tilde{E}_{d} -i\omega _{R} \chi \right)-\frac{\partial }{\partial \chi } \left(-\left(-i\Delta _{ex} +\gamma _{ex} \right)\chi -i\omega _{R} \phi -2i\alpha \chi ^{+} \chi ^{2} \right)+}\\{+\frac{\partial ^{2} }{\partial \chi ^{2} } \left(-i\alpha \chi ^{2} \right)+h.c.]P,}\end{array} \eqno (A.1)
\end{equation*}
where $\chi$ and $\phi$ -- are $c$-numbers.
A solution of the Fokker-Planck equation (A.1) is obtained by the method of potentials in the adiabatic limit \cite{DGG}:
\begin{equation*} \label{EqA2_}
\begin{array}{l} {P_{ss} \left(\chi ,\chi ^{+} \right)=N\chi ^{-2-i\frac{\gamma }{\alpha } } \chi ^{+} {}^{-2+i\frac{\gamma ^{*} }{\alpha } }\exp \left(-\sigma \tilde{E}_{d} \frac{1}{\chi } -\sigma ^{*} \tilde{E}_{d}^{*} \frac{1}{\chi ^{+} } +2\chi \chi ^{+} \right)}\end{array},\eqno (A.2)
\end{equation*}
where $\gamma =\gamma _{ex} +\frac{\omega _{R}^{2} \gamma _{ph} }{\gamma _{ph}^{2} +\Delta _{ph}^{2} } -i\left(\Delta _{ex} +\frac{\Delta _{ph} \omega _{R}^{2} }{\gamma _{ph}^{2} +\Delta _{ph}^{2} } \right)$, $\sigma =\omega _{R} \frac{\gamma _{ph} -i\Delta _{ph} }{\alpha \left(\gamma _{ph}^{2} +\Delta _{ph}^{2} \right)} $. $N$ -- is normalized constant.

We use solution (A.2) to calculate the correlation functions of any order for excitons:
\begin{equation*} \label{EqA3_}
\begin{array}{l} {G^{\left({\rm mn}\right)} =\left\langle \left(\chi ^{+} \right)^{{\rm m}} \chi ^{{\rm n}} \right\rangle =\int \chi ^{+} {}^{m} \chi ^{n} P_{ss}d\mu  =} \\ {(-1)^{n+m}\frac{\left(\sigma \tilde{E}_{{\rm d}} \right)^{{\rm n}} \left(\sigma ^{*} \tilde{E}_{{\rm d}}^{*} \right)^{{\rm m}} \Gamma \left({{\rm i}\gamma ^{*} \mathord{\left/ {\vphantom {{\rm i}\gamma ^{*}  \alpha }} \right. \kern-\nulldelimiterspace} \alpha } \right)\Gamma \left(-{{\rm i}\gamma \mathord{\left/ {\vphantom {{\rm i}\gamma  \alpha }} \right. \kern-\nulldelimiterspace} \alpha } \right)}{\Gamma \left({\rm m}+{{\rm i}\gamma ^{*} \mathord{\left/ {\vphantom {{\rm i}\gamma ^{*}  \alpha }} \right. \kern-\nulldelimiterspace} \alpha } \right)\Gamma \left({\rm n}-{{\rm i}\gamma \mathord{\left/ {\vphantom {{\rm i}\gamma  \alpha }} \right. \kern-\nulldelimiterspace} \alpha } \right)}\frac{{}_{0} {\rm F}_{2} \left({\rm m}+{{\rm i}\gamma ^{*} \mathord{\left/ {\vphantom {{\rm i}\gamma ^{*}  \alpha }} \right. \kern-\nulldelimiterspace} \alpha } ,{\rm n}-{{\rm i}\gamma \mathord{\left/ {\vphantom {{\rm i}\gamma  \alpha }} \right. \kern-\nulldelimiterspace} \alpha } ,2 \left|\sigma \right|^{2} \left|\tilde{E}_{{\rm d}} \right|^{2} \right)}{{}_{0} {\rm F}_{2} \left({{\rm i}\gamma ^{*} \mathord{\left/ {\vphantom {{\rm i}\gamma ^{*}  \alpha }} \right. \kern-\nulldelimiterspace} \alpha } ,-{{\rm i}\gamma \mathord{\left/ {\vphantom {{\rm i}\gamma  \alpha }} \right. \kern-\nulldelimiterspace} \alpha } ,2  \left|\sigma \right|^{2} \left|\tilde{E}_{{\rm d}} \right|^{2} \right)} }
\end{array},\eqno (A.3)
\end{equation*}
where $\Gamma $ is a gamma function and ${}_{0} {\rm F}_{2} $ is a hypergeometric function.

The stochastic differential equations can be obtained in the Ito calculus by converting the Fokker-Planck equation (A.1) into the Ito form \cite{DGG}:
\begin{equation*} \label{EqA4_}
\left\{\begin{array}{l} {\frac{\partial }{\partial t} \phi =-\left(i\Delta _{ph} +\gamma _{ph} \right)\phi +\tilde{E}_{d} -i\omega _{R} \chi ,} \\ {\frac{\partial }{\partial t} \phi ^{+} =-\left(-i\Delta _{ph} +\gamma _{ph} \right)\phi ^{+} +\tilde{E}_{d}^{*} +i\omega _{R} \chi ^{+} ,} \\ {\frac{\partial }{\partial t} \chi =-\left(-i\Delta _{ex} +\gamma _{ex} \right)\chi -i\omega _{R} \phi -2i\alpha \chi ^{+} \chi ^{2}+\left(1-i\right)\sqrt{\alpha } \chi \xi \left(t\right),} \\ {\frac{\partial }{\partial t} \chi ^{+} =-\left(i\Delta _{ex} +\gamma _{ex} \right)\chi ^{+} +i\omega _{R} \phi ^{+} +2i\alpha\chi ^{+2} \chi+\left(1+i\right)\sqrt{\alpha } \chi ^{+} \xi ^{+} \left(t\right),} \end{array}\right.\eqno (A.4)
\end{equation*}
where $\xi \left(t\right)$ is an independent stochastic function, whose correlation functions satisfy the following relations: $\left\langle \xi \left(t\right)\right\rangle =0$, $\left\langle \xi ^{+} \left(t\right)\right\rangle =0$, $\left\langle \xi \left(t\right)\xi ^{+} \left(t'\right)\right\rangle =\delta \left(t-t'\right)$.
\section*{Appendix B:~Derivation of the Wigner function}

\setcounter{equation}{0}

%
We use the Wigner function for a visual presentation of the statistical properties of the exciton-polariton system based on the steady-state solution of the P-function (A.2).
The Wigner function one can be expressed in terms of the P-representation as follow \cite{Kheruntsyan}:

\begin{equation*} \label{EqB1_}
\begin{array}{l} {W\left(\chi\right)=\frac{2}{\pi } e^{-2\left|\chi\right|^{2} } \int _{C_{x } }\int _{C_{x ^{+} } }P_{SS}\left(x ,x ^{+} \right)\exp \left(2\chi^{*} x +2\chi x^{+} -2x ^{+} x \right)d x^{+} dx}\end{array}.\eqno (B.1)
\end{equation*}

Then substituting (A.2) in (B.1) and change variables in the integration $x\rightarrow -E_{d}\sigma/t$, further using the Schlaflis Integral \cite{C}

\begin{equation*} \label{EqB2_}
J_{\nu } \left(z\right)=\frac{\left(z/2\right)^{\nu } }{2\pi i} \int _{C}t^{-\nu -1} \exp \left[t-\frac{z^{2} }{4t} \right]dt,\eqno (B.2)
\end{equation*}
we obtain the follow relation for the Wigner function for the excitons
\begin{equation*} \label{EqB3_}
W\left(\chi \right)=N'e^{-2\left|\chi \right|^{2} } \left|\frac{J_{-i\frac{\gamma }{\alpha } -1} \left(\sqrt{8 \tilde{E}_{d} \sigma \chi^{*}} \right)}{\left(\chi^{*}\right)^{-\left(i\frac{\gamma }{\alpha } +1\right)/2} } \right|^{2}, \eqno (B.3)
\end{equation*}
where $N'$ is a normalization constant, defined by the following expression:
\begin{equation*} \label{EqB4_}
N'=\frac{2}{\pi \left|\left(2 \tilde{E}_{d} \sigma \right)^{-i\frac{\gamma }{\alpha } -1} \right|} \frac{\Gamma \left(-i\frac{\gamma }{\alpha } \right)\Gamma \left(i\frac{\gamma ^{*} }{\alpha } \right)}{{}_{0} {\rm F}_{2} \left({{\rm i}\gamma ^{*} \mathord{\left/ {\vphantom {{\rm i}\gamma ^{*}  \alpha }} \right. \kern-\nulldelimiterspace} \alpha } ,-{{\rm i}\gamma \mathord{\left/ {\vphantom {{\rm i}\gamma  \alpha }} \right. \kern-\nulldelimiterspace} \alpha } ,2\left|\sigma \tilde{E}_{d} \right|^{2} \right)}.\eqno (B.4)
\end{equation*}
Here we use the following expansion of the power series of the Bessel function \cite{Watson}
\begin{equation*} \label{EqB5_}
\begin{array}{l} J_{\nu}(\sqrt{t})=\left(\frac{t}{4}\right)^{\nu/2}\sum\limits_{k=0}^{\infty}\frac{(-1)^{k}(t/4)^{k}}{k!\Gamma(k+\nu+1)}\end{array}.\eqno (B.5)
\end{equation*}

As you can see, the Wigner function is always positive and is and contains the square module of the Bessel function with a complex index $-i\gamma/\alpha-1$.

We use the governing principle \eqref{Eq3_} for the transition to the Wigner function of photons:

\begin{equation*} \label{EqB6_}
\begin{array}{l} {W\left(\phi \right)=N''e^{-2\frac{\left(\gamma _{ph} Im\phi +\Delta _{ph} Re\phi \right)^{2} +\left(\gamma _{ph} Re\phi -\Delta _{ph} Im\phi -\tilde{E}_{d} \right)^{2} }{\omega _{R}^{2} } } \times } \\ {\times \left|\frac{J_{-i\frac{\gamma }{\alpha } -1} \left(\sqrt{8\tilde{E}_{d} \sigma \frac{\left(i\tilde{E}_{d} +\phi *\left(\Delta _{ph} -i\gamma _{ph} \right)\right)}{\omega _{R} } } \right)}{\left(\frac{\left(i\tilde{E}_{d} +\phi *\left(\Delta _{ph} -i\gamma _{ph} \right)\right)}{\omega _{R} } \right)^{\left(-i\frac{\gamma }{\alpha } -1\right)/2} } \right|^{2}}\end{array},\eqno (B.5)
\end{equation*}
where, we introduced a new normalized constant $ N''=\frac{\omega _{R}^{2} }{\Delta _{ph}^{2} +\gamma _{ph}^{2} } N'.$

\end{document}